\documentclass[
reprint,
amsmath,
amssymb,
aip,
superscriptaddress,
notitlepage
]{revtex4-2}

\usepackage[colorlinks, breaklinks=true, linkcolor=blue, citecolor=blue, linktocpage=true]{hyperref}
\usepackage{color}
\usepackage{graphicx}
\usepackage{dcolumn}
\usepackage{bm}

\usepackage{txfonts}

\begin{document}

\title{Emergent One-Dimensional Helical Channel in Higher-Order Topological Insulators with Step Edges}

\author{Akihiko Sekine}
\email{akihiko.sekine@fujitsu.com}
\affiliation{Fujitsu Research, Fujitsu Limited, Atsugi, Kanagawa 243-0197, Japan}
\author{Manabu Ohtomo}
\affiliation{Fujitsu Research, Fujitsu Limited, Atsugi, Kanagawa 243-0197, Japan}
\author{Kenichi Kawaguchi}
\affiliation{Fujitsu Research, Fujitsu Limited, Atsugi, Kanagawa 243-0197, Japan}
\author{Mari Ohfuchi}
\affiliation{Fujitsu Research, Fujitsu Limited, Atsugi, Kanagawa 243-0197, Japan}

\date{\today}

\begin{abstract}
We study theoretically the electronic structure of three-dimensional (3D) higher-order topological insulators in the presence of step edges.
We numerically find that a 1D conducting state with a helical spin structure, which also has a linear dispersion near the zero energy, emerges at a step edge and on the opposite surface of the step edge.
We also find that the 1D helical conducting state on the opposite surface of a step edge emerges when the electron hopping in the direction perpendicular to the step is weak.
In other words, the existence of the 1D helical conducting state on the opposite surface of a step edge can be understood by considering an addition of two different-sized independent blocks of 3D higher-order topological insulators.
On the other hand, when the electron hopping in the direction perpendicular to the step is strong, the location of the emergent 1D helical conducting state moves from the opposite surface of a step edge to the dip ($270^{\circ}$ edge) just below the step edge.
In this case, the existence at the dip below the step edge can be understood by assigning each surface with a sign ($+$ or $-$) of the mass of the surface Dirac fermions.
These two physical pictures are connected continuously without the bulk bandgap closing.
Our finding paves the way for on-demand creation of 1D helical conducting states from 3D higher-order topological insulators employing experimental processes commonly used in thin-film devices, which could lead to, e.g., a realization of high-density Majorana qubits.
\\
\\
\end{abstract}

\maketitle

\section{Introduction}
Since the discovery of topological insulators, topologically nontrivial phases of matter have attracted broad attention, not only from the viewpoint of fundamental research but also from the viewpoint of possible technological applications.
Regarding the latter viewpoint, for example, realizing Majorana zero modes enables topological quantum computation \cite{Nayak2008,Alicea2012,Sarma2015,Sato2016}, which has low error rate and whose qubits are robust against noises due to the topological nature of Majorana fermions.
Also, possible ways to manipulate and utilize the spin-momentum locked helical surface states of three-dimensional (3D) topological insulators have been investigated experimentally in spintronics \cite{Mellnik2014,Li2014,Fan2014,Shiomi2014,Fan2016,He2022}.

Higher-order topological insulators are a new class of topological materials, which are generalization of ``conventional'' topological insulators and are characterized as insulators that have topological localized states at least two dimensions lower than the bulk \cite{Benalcazar2017,Song2017,Langbehn2017,Benalcazar2017a,Schindler2018,Ezawa2018}.
Namely, 2D (3D) second-order topological insulators have topological 0D ``corner'' (1D ``hinge'') states.
Third-order topological insulators exists only in three spatial dimensions, having topological 0D corner states.
The focus of this paper is 3D second-order topological insulators, whose experimental signatures have been observed in Bismuth \cite{Schindler2018a,Aggarwal2021}, WTe$_2$ \cite{Choi2020,Kononov2020,Huang2020}, and Bi$_4$Br$_4$ \cite{Noguchi2021,Shumiya2022}.

Realizing a high-quality edge is an important issue for practical applications of 1D helical conducting channels localized at the edges of a sample, such as the edge states of 2D topological insulators and the hinge states of 3D higher-order topological insulators.
For example, oxidization of the samples (and therefore their edges or hinges) can cause the degradation of the transport properties such as conductivity \cite{Ye2016,Woods2017,Hou2020}, and hence is one of the problems that should be avoided.
Also, in the case of 3D higher-order topological insulators, the appearance of a 1D helical channel depends on the direction of the edge with respect to the crystal axes, implying that the straightness of the edge is an important factor for obtaining a high conductivity.
As a possible route to avoid such potential issues, in this paper we propose an alternative way of realizing 1D helical conducting states on the 2D surface of 3D higher-order topological insulators by creating step edges.

In this paper, we theoretically study the electronic structure of 3D higher-order (second-order) topological insulators in the presence of step edges.
From the diagonalization of a tight-binding model for few-layer Td-$X$Te$_2$ ($X$ = Mo, W) in a 3D geometry with a step edge (see Fig.~\ref{Fig1}), we find that a 1D conducting state with a helical spin structure, which also has a linear dispersion near the zero energy, emerges at the step edge and on the opposite surface of the step edge.
By varying the parameters in the tight-binding Hamiltonian, we also find that the 1D helical conducting state on the opposite surface of a step edge emerges when the electron hopping in the direction perpendicular to the step is weak.
Such an existence of the emergent 1D helical conducting state on the opposite surface of a step edge in the weak coupling regime can be understood by considering an addition of different-sized independent blocks of 3D higher-order topological insulators.
On the other hand, when the electron hopping in the direction perpendicular to the step is strong, it is found that the location of the emergent 1D helical conducting state moves from the opposite surface of a step edge to the dip ($270^{\circ}$ edge) just below the step edge.
In this case, the existence at the dip below the step edge can be understood by assigning each surface with a sign ($+$ or $-$) of the mass of the surface Dirac fermions.
We propose a possible experimental setup for creating 1D helical conducting channels as desired by controlling the locations and the numbers of step edges in 3D higher-order topological insulators, which utilizes experimental processes commonly used in thin-film devices.
We expect that our finding enables on-demand creation of 1D helical conducting states utilizing 3D higher-order topological insulators.

This paper is organized as follows.
In Sec.~\ref{Sec-Model} we present an effective model of few-layer Td-$X$Te$_2$ ($X$ = Mo, W) describing a 3D higher-order topological insulator phase, and a calculation method for obtaining the energy spectrum and wave function of the system in the geometry with a step edge.
In Sec.~\ref{Sec-NumericalResults} we show the emergence of a 1D helical conducting state at the step edge and on the opposite surface of the step edge, using the parameters intended for WTe$_2$.
In Sec.~\ref{Sec-Mechanism} we attempt to understand how and when a 1D helical conducting state on the opposite surface of the step edge emerges by varying the geometry of the system and varying the parameters in the Hamiltonian continuously without the bulk bandgap closing (i.e., without the change in the bulk topology).
In Sec.~\ref{Sec-Discussion} we discuss a possible application of our finding and a relevance of our finding  to a recent experiment in few-layer WTe$_2$.
We also discuss the difference between our study and earlier theoretical and experimental studies showing the presence of 1D helical conducting states at step edges.
In Sec.~\ref{Sec-Experimental} we discuss a possible experimental setup for the realization and observation of the emergent 1D helical conducting state on the opposite surface of a step edge.
We also propose a possible setup for on-demand creation of multiple Majorana fermions on a 2D surface of a single material.
Finally, in Sec.~\ref{Sec-Summary} we summarize this study.

\section{Theoretical Model \label{Sec-Model}}
\begin{figure}[!t]
\centering
\includegraphics[width=\columnwidth]{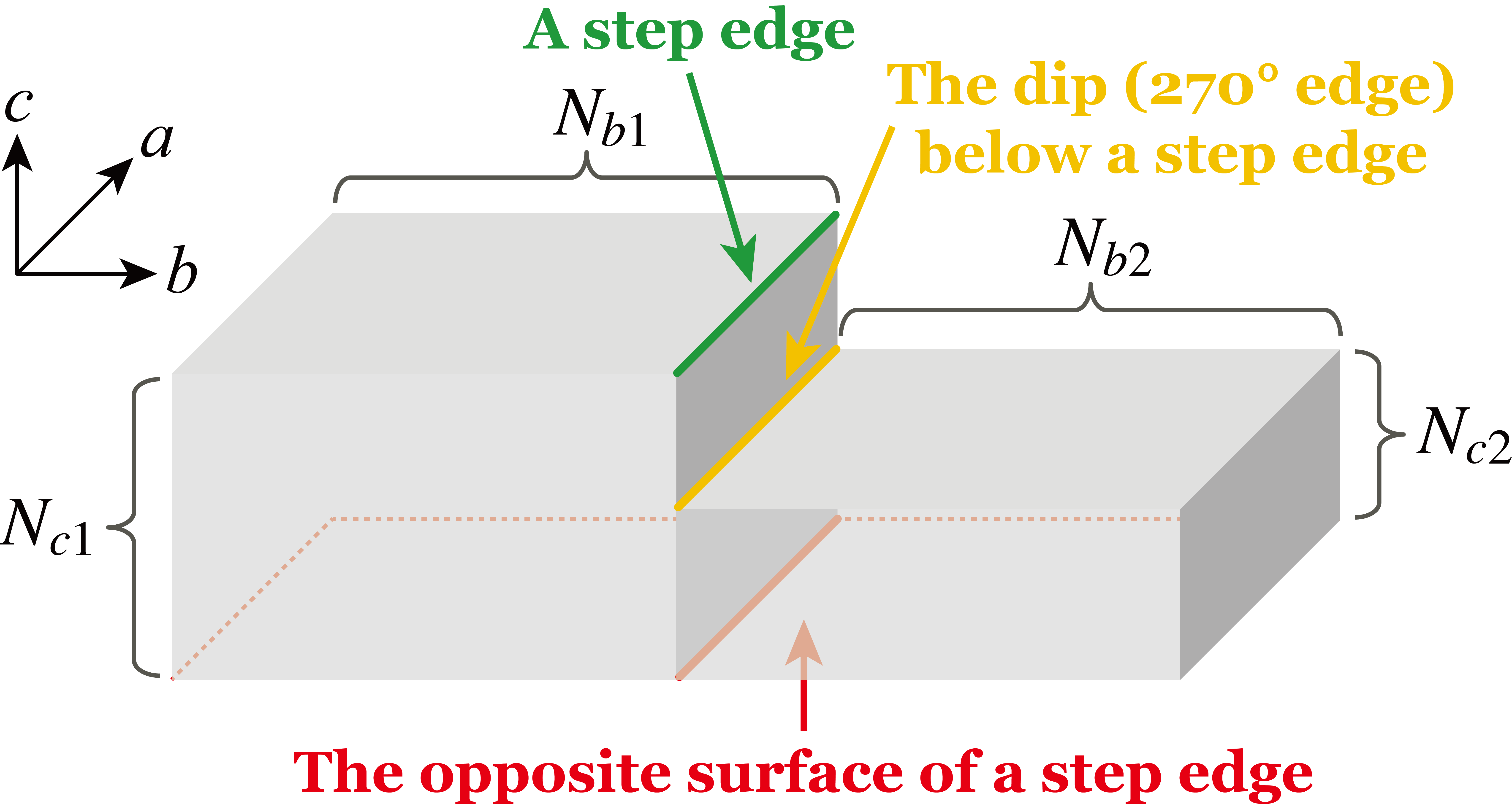}
\caption{Schematic illustration of the geometry with a step edge along the $a$ direction.
$N_{b1}$ and $N_{b2}$ ($N_{c1}$ and $N_{c2}$) are the numbers of sites in the $b$ ($c$) direction.
In our model, a 1D helical channel emerges (denoted by a red solid line) on the surface opposite to a step edge when $N_{c1}>N_{c2}$.
}\label{Fig1}
\end{figure}
\begin{figure*}[!t]
\centering
\includegraphics[width=2\columnwidth]{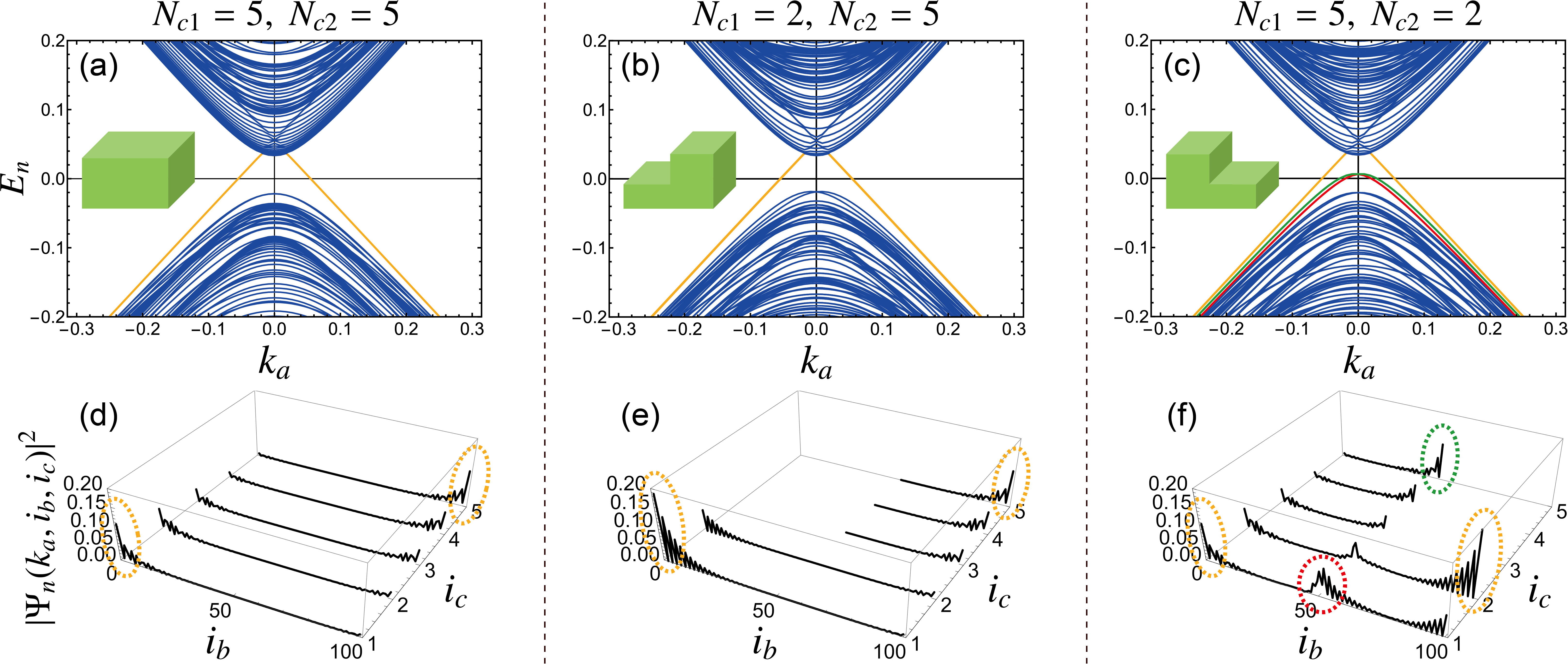}
\caption{(a) through (c) [(d) through (f)] The energy spectrum (site-resolved probability distribution with $k_a=0.05$) of the system without a step, with a positive-slope step, and with a negative-slope step, respectively.
The bulk bands are displayed blue.
The bands of the 1D localized states are highlighted in yellow, green, and red.
Note that each band in (b) and (c) is not doubly degenerate due to the inversion symmetry breaking, although the bands of the ordinary topological hinge states (yellow) look almost degenerate.
In (d) through (f), the locations of the 1D localized states are highlighted by dashed lines of the corresponding colors to the energy bands.
The parameters are set to be $N_{b1}=N_{b2}=50$, $m_1=-3$, $m_2=0.3$, $m_3=0.2$, $v_a=2$, $v_b=1.6$, $v_c=1$, $\lambda_b=0.1$, $\lambda_c=1$, $\gamma_x=0.4$, $\gamma_z=-0.4$, and $\beta_a=1.5$.
}\label{Fig2}
\end{figure*}
As a concrete example of 3D higher-order topological insulators, we consider the low-energy effective model of few-layer Td-$X$Te$_2$ ($X$ = Mo, W) \cite{Wang2019,Ezawa2019,Choi2020}.
The bulk Td-$X$Te$_2$ is a layered transition-metal dichalcogenide and is considered a Type-II Weyl semimetal with broken inversion symmetry \cite{Soluyanov2015,Sun2015, Wang2016,Li2017}.
However, in the few-layer limit, it has been suggested that the Weyl points are annihilated under realistic conditions such as symmetry-preserving strain or lattice distortion, and that the system becomes a nonsymmetry-indicated, noncentrosymmetric 3D higher-order (second-order) topological insulator \cite{Wang2019,Choi2020}.

Although Td-$X$Te$_2$ has noncentrosymmetric $C_{2v}$ point group symmetry, Td-$X$Te$_2$ can be treated as having the same topological property as 1T$^\prime$-$X$Te$_2$ with centrosymmetric $C_{2h}$ point group symmetry, when the Weyl points are annihilated \cite{Wang2019,Choi2020}.
Accordingly, the low-energy effective Hamiltonian describing the higher-order topological insulator phase of few-layer Td-$X$Te$_2$ possesses time-reversal symmetry, inversion symmetry, two-fold rotational symmetry around the $a$ axis, and mirror symmetry with the mirror plane perpendicular to the rotational axis.
The low-energy effective Hamiltonian on a cubic lattice is given by \cite{Wang2019,Choi2020}
\begin{align}
\mathcal{H}_{\mathrm{eff}}(\bm{k})=&\ \left(m_1+\sum_{j=a,b,c}v_j\cos k_j+m_2\mu_x+m_3\mu_z\right)\tau_z\nonumber\\
&+\lambda_b\sin k_b\mu_y\tau_y+\lambda_c\sin k_c\tau_x+\gamma_x\mu_x+\gamma_z\mu_z\nonumber\\
&+\beta_a\sin k_a\mu_z\tau_y\sigma_z,
\label{Effective-Hamiltonian-continuum}
\end{align}
where $\bm{k}=(k_a,k_b,k_c)$ is a wave vector, $\sigma_i$ are the Pauli matrices acting on the spin space, and $\tau_i$ and $\mu_i$ are the Pauli matrices acting on the $(s, p)$ orbital space and on the $(s_1, s_2)$, $(ip_1, ip_2)$ orbital space, respectively.
The $c$ axis is the stacking direction.
The Hamiltonian~(\ref{Effective-Hamiltonian-continuum}) possesses time-reversal symmetry $\mathcal{T}=i\tau_z\sigma_y K$, inversion symmetry $\mathcal{P}=\tau_z$, two-fold rotational symmetry $C_{2a}=i\tau_z\sigma_y$ around the $a$ axis, and mirror symmetry $\mathcal{M}_a=\mathcal{P}C_{2a}=i\sigma_y$ with the mirror plane perpendicular to the $C_{2a}$ rotational axis.
Therefore, the Hamiltonian possesses the $C_{2h}$ point group symmetry which is the same symmetry as the point group symmetry of 1T$^\prime$-WTe$_2$.
The Hamiltonian~(\ref{Effective-Hamiltonian-continuum}) has 1D helical hinge states along the $a$ axis [see Figs.~\ref{Fig2}(a) and \ref{Fig2}(d)].

To see how the presence of a step edge along the $a$ direction affects the electronic structure of the system, let us consider a real-space version of Eq.~(\ref{Effective-Hamiltonian-continuum}) in the tight-binding approximation with only the nearest-neighbor hopping taken into account in a geometry depicted in Fig.~\ref{Fig1}. 
We use periodic boundary condition in the $a$ direction and open boundary condition in the $b$ and $c$ directions.
Then, the resulting Hamiltonian reads
\begin{align}
H_{\mathrm{eff}}=\sum_{k_a}c^\dag_{k_a}H(k_a)c_{k_a},
\label{Lattice-Hamiltonian}
\end{align}
where $c_{k_a}$ is an $8(N_{b1}N_{c1}+N_{b2}N_{c2})$-component electron annihilation operator, with $N_{b1}$ and $N_{b2}$ ($N_{c1}$ and $N_{c2}$) being the numbers of sites in the $b$ ($c$) direction.
The energy spectrum $E_n(k_a)$ and wave function $\Psi_n(k_a)$ of the system described by the Hamiltonian~(\ref{Lattice-Hamiltonian}) is obtained by diagonalizing $H(k_a)$ for each momentum $k_a$.
Namely, the relation $H(k_a)\Psi_n(k_a)=E_n(k_a)\Psi_n(k_a)$ is satisfied.

\section{Numerical Results \label{Sec-NumericalResults}}
\begin{figure*}[!t]
\centering
\includegraphics[width=2\columnwidth]{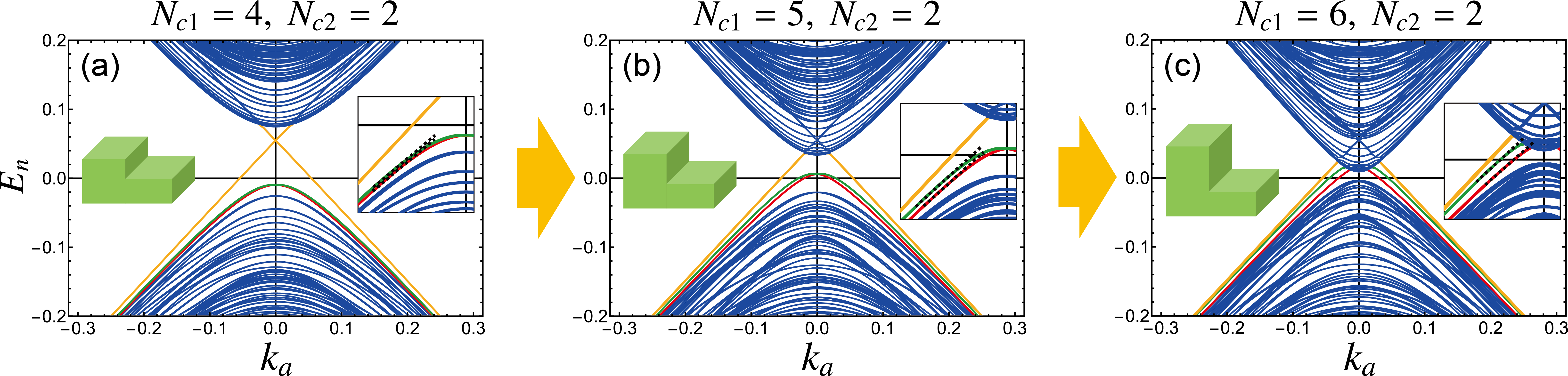}
\caption{(a) through (c) Evolution of the energy spectrum of the system with a negative-slope step with respect to the increasing value of $N_{c1}$ from $4$ to $6$.
The value of $N_{c2}$ is fixed to $2$.
The insets highlight (by black dashed lines) the linear dependence of the emergent 1D states on momentum $k_a$ near the zero-energy level $E_n=0$.
The other parameters are the same as in Fig.~\ref{Fig2}.
}\label{Fig3}
\end{figure*}
In this section, we show the numerical results obtained by using the parameters intended for WTe$_2$ \cite{Choi2020}.
For clarity, let us define a ``positive-slope (negative-slope) step'', in which the height of the left part $N_{c1}$ is smaller (larger) than that of the right part $N_{c2}$.

First, for reference we show the energy spectrum and probability distribution for the system without a step (i.e., in a rectangular geometry) in Figs.~\ref{Fig2}(a) and \ref{Fig2}(d), respectively, from which we see the presence of the ordinary 1D topological hinge states.
Next, we show in Figs.~\ref{Fig2}(b) and \ref{Fig2}(e) [Figs.~\ref{Fig2}(c) and \ref{Fig2}(f)] the energy spectrum and probability distribution of the system with a positive-slope step (a negative-slope step), respectively.
We find that, in addition to the ordinary topological hinge states, 1D in-gap localized states emerge at the step edge and on the opposite surface of the step edge in the case of a negative-slope step, while they do not emerge in the case of a positive-slope step.
We shall discuss in detail the physical mechanism for the emergence or absence of these 1D in-gap localized states in Sec.~\ref{Sec-Understanding}.
We stress here that in all the three geometries (one without a step edge, with a positive-slope step, or with a negative-slope step) the system belongs to the same second-order topological insulator phase, because the bulk bandgap does not close when changing the geometry of the system. 

Figure~\ref{Fig3} shows the evolution of the energy spectrum of the system with a negative-slope step with respect to the increasing difference between the number of layers, $N_{c1}-N_{c2}$.
It can be seen that, as the difference between the number of layers, $N_{c1}-N_{c2}$, becomes larger, the dispersion of the 1D states emerging at the step edge and on the opposite surface of the step edge moves toward the bulk conduction bands.
We also find that the region of the dispersion that has the linear dependence on momentum $k_a$ (i.e., $E_n\propto k_a$) begins to intersect the zero-energy $E_n=0$ line and it becomes more distinguishable from the bulk valence bands, as the value of $N_{c1}-N_{c2}$ which corresponds to the height of the step becomes larger.
We have also checked what would happen when the value of $N_{c1}$ is fixed while the value of $N_{c2}$ is varied.
Starting from a small step height (like $N_{c1}=6$ and $N_{c2}=5$), as the step height becomes larger, their dispersion begins to intersect the zero-energy line and to be linear in momentum as well as becoming more distinguishable from the bulk conduction bands.
This result indicates that the step height is an important factor in determining their dispersion.
In Sec.~\ref{Sec-Parameter} we will argue that such a linearity and gaplessness feature of the 1D localized states at the step edge and on the opposite surface of the step edge in the case of $N_{c1}\gg N_{c2}$ is ascribed to that their origin is the ordinary 1D topological hinge state.
We will also argue that their gapped dispersion is due to the finite-size effect in the $c$ direction, i.e., due to the overlap between their wave functions.
As is readily seen from Fig.~\ref{Fig3}, the magnitude of the bulk bandgap is determined by the number of layers $N_{c1}$, which is larger than $N_{c2}$.
The same argument also applies to the case of positive-slope steps with $N_{c1}<N_{c2}$.
Accordingly, the bulk bandgap closes when $N_{c1}\geq 7$ or $N_{c2}\geq 7$.

\begin{figure}[!t]
\centering
\includegraphics[width=\columnwidth]{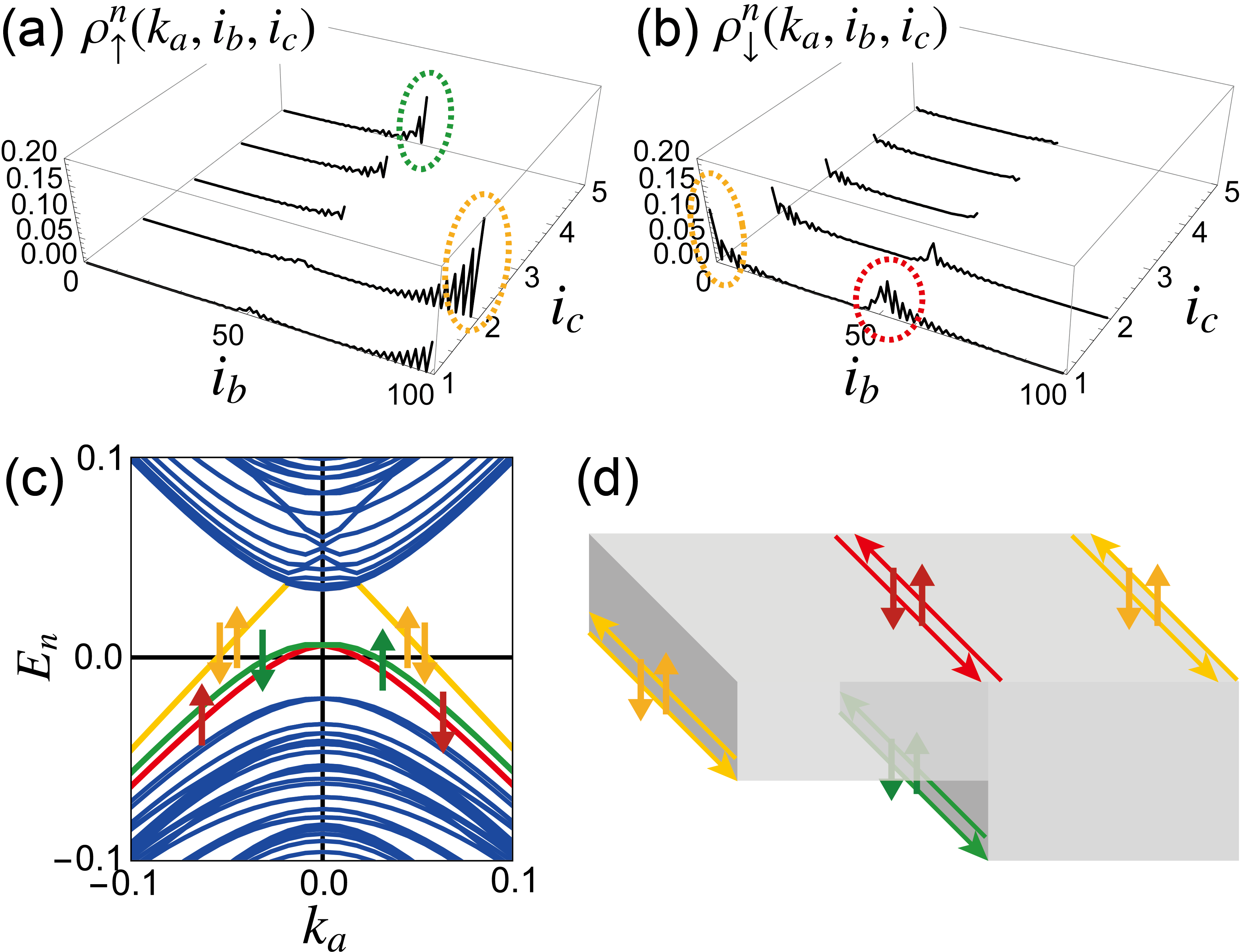}
\caption{(a) and (b) The up-spin (down-spin) component of the electron density [Eq.~(\ref{spin-density})] with $k_a=0.05$.
(c) Helical spin structure of the 1D localized states.
(d) Schematic Illustration of the spin-momentum locked 1D localized states.
The system size is $N_{b1}=N_{b2}=50$, $N_{c1}=5$, and $N_{c2}=2$.
The other parameters are the same as in Fig.~\ref{Fig2}.
}\label{Fig4}
\end{figure}
Next, we investigate the spin structure of the 1D conducting states emerging at the step edge and on the opposite surface of the step edge.
The electron density of a given momentum $k_a$ and spin $\sigma=\uparrow,\downarrow$ at site $(i_b, i_c)$ is defined by
\begin{align}
\rho^n_{\sigma}(k_a,i_b,i_c)=\sum_{\mu_z=\pm1}\sum_{\tau_z=\pm1}|\Psi^n_{\sigma}(k_a,i_b,i_c)|^2,
\label{spin-density}
\end{align}
where $\Psi^n_{\sigma}(k_a,i_b,i_c)$ is the site-resolved wave function of band $n$  with spin $\sigma$, and $\sum_{\mu_z=\pm1}$ and $\sum_{\tau_z=\pm1}$ indicate the summations over the orbital components of the wave function.
Here, the wave function is normalized as $\sum_{i_b,i_c}[\rho^n_{\uparrow}(k_a,i_b,i_c)+\rho^n_{\downarrow}(k_a,i_b,i_c)]=1$.
The electron density with spin-up $\uparrow$ and spin-down $\downarrow$ are shown in Figs.~\ref{Fig4}(a) and \ref{Fig4}(b), respectively.
We find that the electrons' spin of the emergent 1D states is fully polarized, as in the case of the ordinary 1D topological hinge states.
In other words, the spin structure of the emergent 1D states is also spin-momentum locked, i.e., helical.
Figures~\ref{Fig4}(c) and \ref{Fig4}(d) illustrate the helical spin structure of the 1D states including the ordinary hinge states in momentum space and real space, respectively.
Note that this helical spin structure can be understood by the fact that the system we consider has time-reversal symmetry.

\section{Physical Mechanism for the Emergence of the 1D Helical Conducting State \label{Sec-Mechanism}}
In this section, we attempt to understand how and when a 1D helical conducting state on the opposite surface of a step edge emerges by varying the geometry of the system and varying the parameters in the Hamiltonian continuously without the bulk bandgap closing (i.e., without the change in the bulk topology).
In Sec.~\ref{Sec-Continuous-Deformation} we consider a continuous deformation of the system geometry to see the origin of the the emergent 1D helical conducting states at the step edge and on the opposite surface of the step edge.
In Sec.~\ref{Sec-Parameter} we investigate the dependence of the electronic structure on the parameters in the Hamiltonian~(\ref{Effective-Hamiltonian-continuum}) to find the condition for the emergence of the 1D helical conducting state on the opposite surface of the step edge.
In Sec.~\ref{Sec-Understanding} we discuss the physical mechanism for the emergence.

\begin{figure*}[!t]
\centering
\includegraphics[width=2\columnwidth]{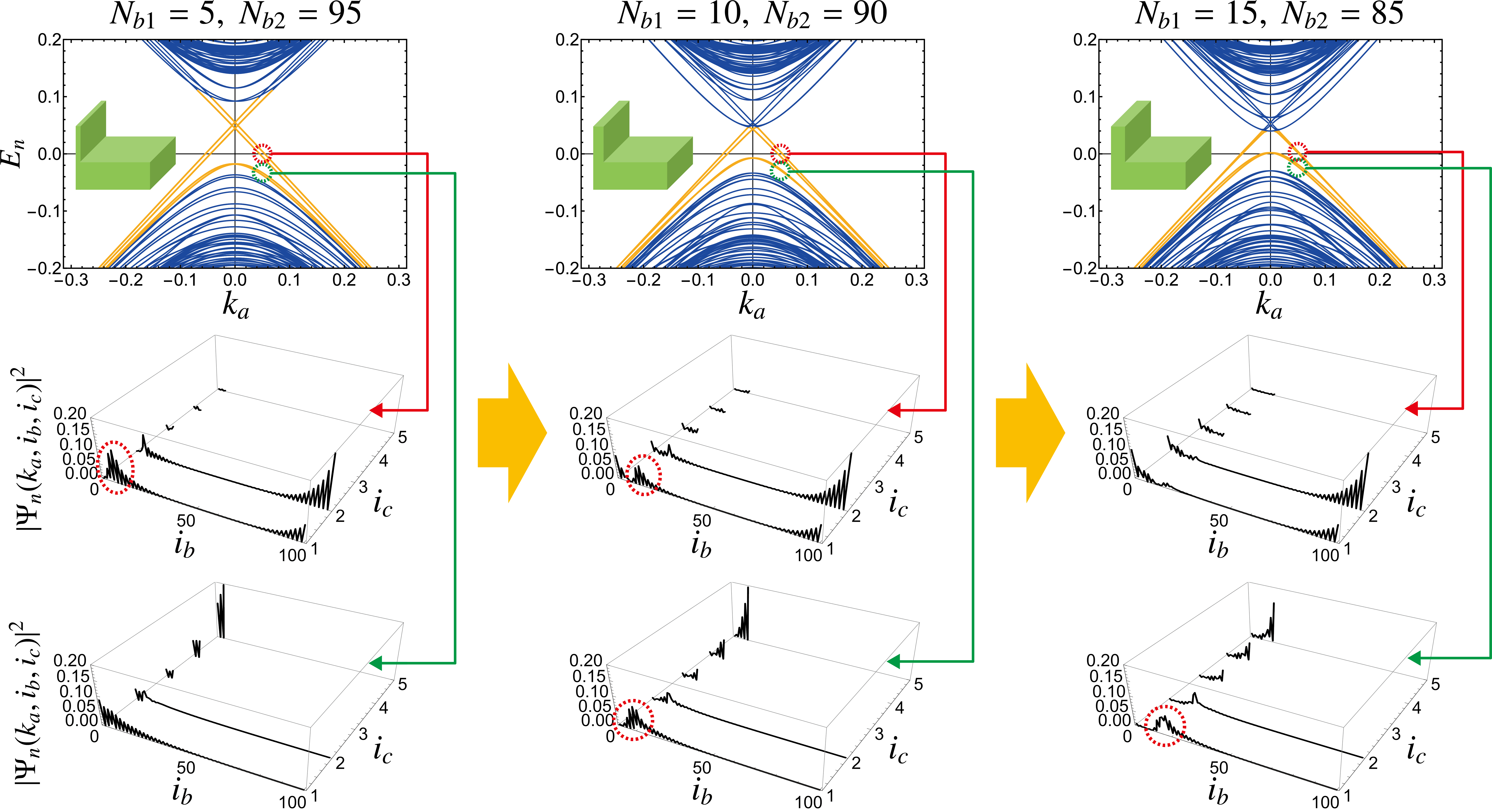}
\caption{
Evolution of the energy spectrum and probability distribution with $k_a=0.05$ under a continuous deformation of the geometry.
From left to right: The width of the left-hand-side block, $N_{b1}$, is varied from $N_{b1}=5$ to $N_{b1}=15$ while the width of the right-hand-side block, $N_{b2}$, is varied under the condition such that $N_{b1}+N_{b2}=100$.
The 1D localized state whose dispersion is gapless (gapped) inside the bulk bandgap is directed by red (green) arrows to the corresponding probability distribution.
Red dashed circles in the probability distributions highlight the 1D localized state on the opposite surface of the step edge.
We set $N_{c1}=5$ and $N_{c2}=2$.
The other parameters are the same as in Fig.~\ref{Fig2}.
}\label{Fig5}
\end{figure*}
\subsection{Continuous deformation of the geometry with $N_{b1}\ll N_{b2}$ \label{Sec-Continuous-Deformation}}
In order to understand the origin of the emergent 1D helical conducting state on the opposite surface of the step edge, we consider the case in which the width of the left-hand-side block is much smaller than that of the right-hand side block (i.e., $N_{b1}\ll N_{b2}$).
Figure~\ref{Fig5} shows the energy spectrum and probability distribution under a continuous deformation of the geometry with $N_{b1}\ll N_{b2}$.
Here, the width of the left-hand-side block, $N_{b1}$, is changed from $N_{b1}=5$ to $N_{b1}=15$, while the width of the right-hand-side block, $N_{b2}$, is changed under the condition such that $N_{b1}+N_{b2}=100$.

When the width of the left-hand-side block is quite small ($N_{b1}=5$; left figures), it can be seen that the dispersion of the emergent 1D helical conducting state on the opposite surface of the step edge is the same as the dispersion of the ordinary topological hinge states, i.e., is linear and gapless (see the red dashed circle in the probability distribution).
This indicates that the system can be viewed as a single block of a higher-order topological insulator with $N_{c2}=2$.
In other words, the presence of the left-hand-side block does not affect largely the electronic structure of the right-hand-side block.
On the other hand, when the width of the left-hand-side block becomes larger ($N_{b1}=15$; right figures), it can be seen that the dispersion of the emergent 1D helical conducting state on the opposite surface of the step edge becomes gapped and the dispersion that is linear and gapless originates from the ordinary 1D topological hinge states at the lower-leftmost and upper-rightmost corners [as we have also seen in Figs.~\ref{Fig2}(c) and \ref{Fig2}(f)].
In between the above two geometries, when the width of the left-hand-side block is intermediate ($N_{b1}=10$; middle figures), an crossover occurs: The state whose dispersion is linear and gapless is localized {\it both} at the lower-leftmost corner and on the opposite surface of the step edge.
In other words, a 1D helical conducting state with a linear and gapless dispersion and the one with a gapped dispersion coexist on the opposite surface of the step edge.

A similar continuous deformation with $N_{b1}\gg N_{b2}$ should be considered in order to understand the origin of the emergent 1D helical conducting state at the step edge (see Appendix~\ref{Appendix-Continuous-Deformation}).
As can be seen in Figs.~\ref{Fig5} and \ref{Fig-Appendix}, the energy dispersion and real-space position of the 1D helical conducting states change continuously without the bulk gap closing.
Recall here that the topological properties of a system does not change under a continuous parameter change without the bulk gap closing.
Hence, combining the results on the continuous deformation in the two limits of $N_{b1}\ll N_{b2}$ and $N_{b1}\gg N_{b2}$, we may conclude that the origin of the emergent 1D helical conducting states on the opposite surface of the step edge and at the step edge are the ordinary topological 1D hinge states.

\subsection{Parameter dependence \label{Sec-Parameter}}
\begin{table*}[!t]
\caption{
Comparison of the Hamiltonian parameters for the two systems we consider.
The parameters for System 1 are used in Fig.~\ref{Fig2}, Fig.\ref{Fig3}, Fig.~\ref{Fig4}, Fig.~\ref{Fig5}, and Fig.~\ref{Fig8}, while those for System 2 are used in Fig.~\ref{Fig6}, Fig.~\ref{Fig7}, and Fig.~\ref{Fig9}.
}
\begin{ruledtabular}
\begin{tabular}{cccccccccccc}
System & $m_1$ & $m_2$ & $m_3$ & $v_a$ & $v_b$ & $v_c$ & $\lambda_b$ & $\lambda_c$ & $\gamma_x$ & $\gamma_z$ & $\beta_a$ \\
\hline
System 1 \cite{Choi2020} & $-3$ & $0.3$ & $0.2$ & $2$ & $1.6$ & $1$ & $0.1$ & $1$ & $0.4$ & $-0.4$ & $1.5$ \\
System 2 \cite{Wang2019} & $-3$ & $0.3$ & $0.2$ & $2$ & $1$ & $1$ & $1$ & $1$ & $0.2$ & $-0.4$ & $1.2$
\end{tabular}
\end{ruledtabular}\label{Table1}
\end{table*}
\begin{figure}[!t]
\centering
\includegraphics[width=\columnwidth]{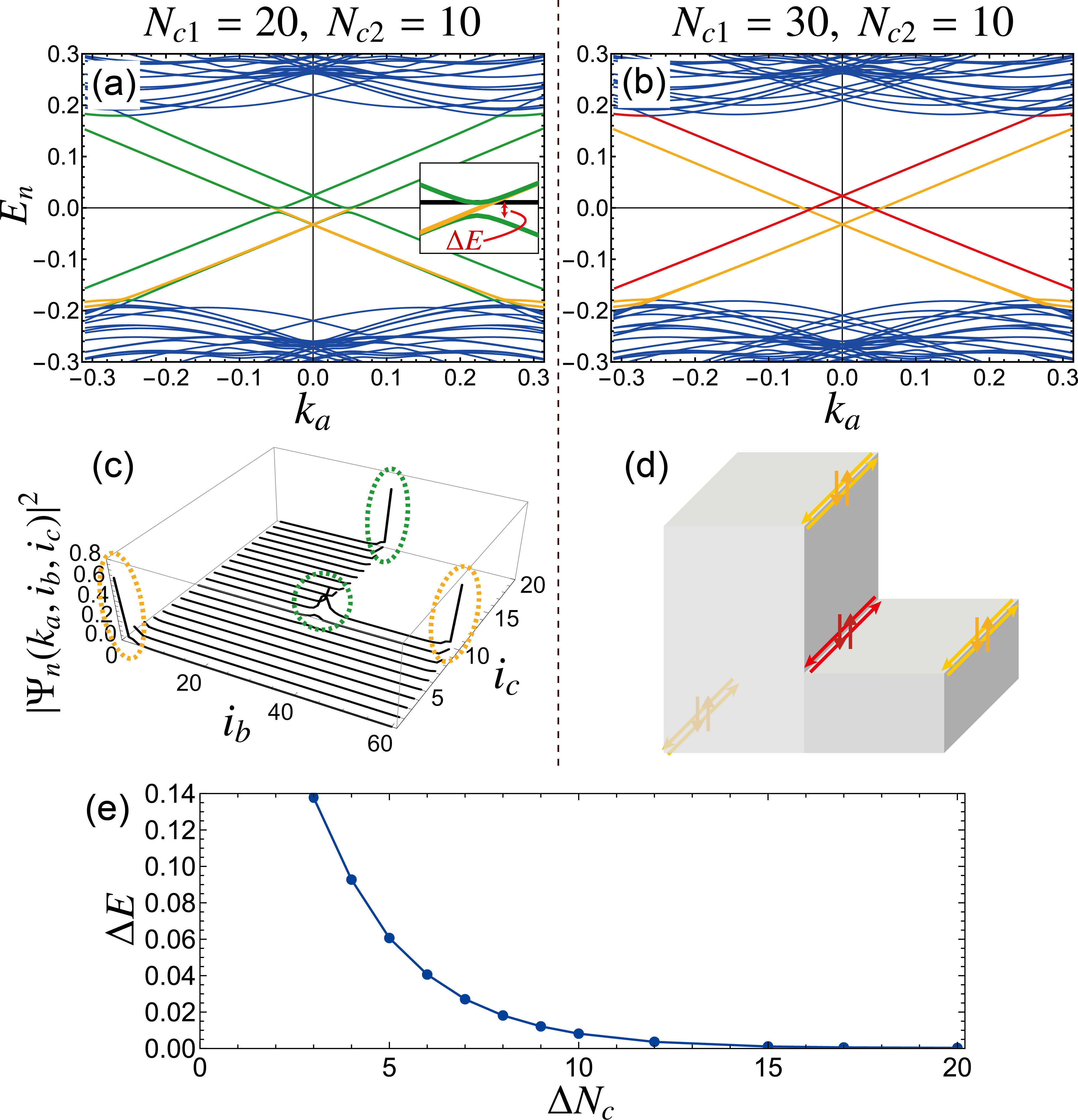}
\caption{(a) The energy spectrum and (c) site-resolved probability distribution with $k_a=0.05$ of a system that has a larger bulk bandgap than the system we have investigated in Sec.~\ref{Sec-NumericalResults}.
The system size is $N_{c1}=20$ and $N_{c2}=10$.
The inset in (a) shows an enlarged view near the gap $\Delta E$ of the energy spectrum of the localized states.
(b) The energy spectrum of a system with the size $N_{c1}=30$ and $N_{c2}=10$.
(d) Schematic illustration of the spin-momentum locked 1D localized states whose energy spectrum is gapless and highlighted in (b) by the same colors.
(e) The energy gap $\Delta E$ as a function of the step height $\Delta N_{c}=N_{c1}-N_{c2}$ at $k_a=0.049$.
The parameters are chosen to be $N_{b1}=N_{b2}=30$, $m_1=-3$, $m_2=0.3$, $m_3=0.2$, $v_a=2$, $v_b=1$, $v_c=1$, $\lambda_b=1$, $\lambda_c=1$, $\gamma_x=0.2$, $\gamma_z=-0.4$, and $\beta_a=1.2$.
}\label{Fig6}
\end{figure}
\begin{figure*}[!t]
\centering
\includegraphics[width=2\columnwidth]{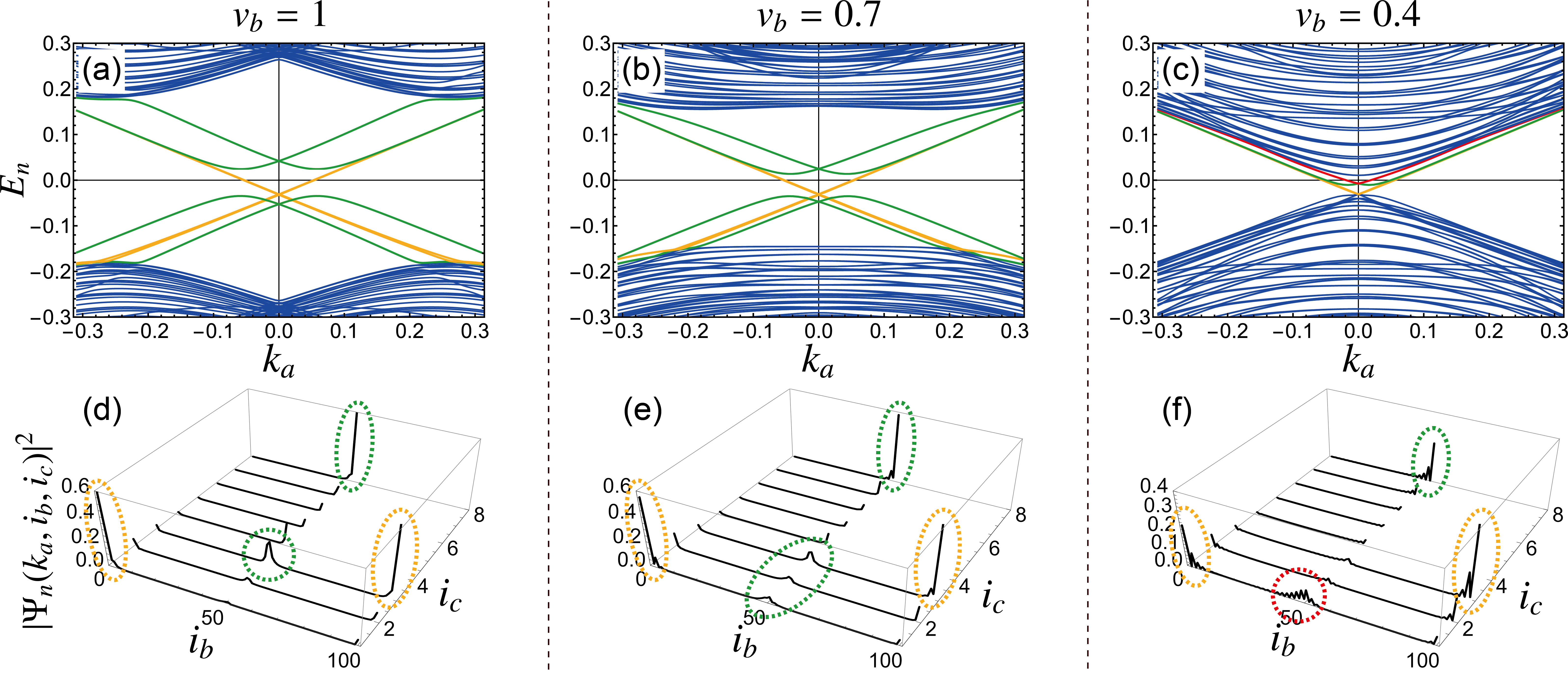}
\caption{
(a) through (c) [(d) through (f)] The energy spectrum (site-resolved probability distribution with $k_a=0.05$) of a system with $v_b=1$, $v_b=0.7$, and $v_b=0.4$, respectively.
The system size is fixed to $N_{b1}=N_{b2}=50$, $N_{c1}=8$, and $N_{c2}=3$.
The other parameters are the same as in Fig.~\ref{Fig6}.
}\label{Fig7}
\end{figure*}
As we have seen in Fig.~\ref{Fig3}, the bulk band gap closes when $N_{c1}\ge 7$ or $N_{c2}\ge 7$ in the calculation intended for WTe$_2$ in Sec.~\ref{Sec-NumericalResults}.
Then, a natural question arises: Is the presence of the 1D helical conducting state on the opposite surface of a step edge stable against (or equivalently, how does it depend on) the increasing values of $N_{c1}$ and $N_{c2}$ in systems whose bulk bandgap is still open even with larger values of $N_{c1}$ and $N_{c2}$?
To shed light on this problem, we use the following model parameters for $X$Te$_2$ ($X$ = Mo, W), which describe a higher-order topological insulator state with a larger bandgap than that we have investigated so far: $m_1=-3$, $m_2=0.3$, $m_3=0.2$, $v_a=2$, $v_b=1$, $v_c=1$, $\lambda_b=1$, $\lambda_c=1$, $\gamma_x=0.2$, $\gamma_z=-0.4$, and $\beta_a=1.2$ \cite{Wang2019}.
See Table~\ref{Table1} for the comparison of the parameters for these two systems.

With these values of the parameters, as shown in Figs.~\ref{Fig6}(a) and \ref{Fig6}(c), we find that 1D helical conducting states appear at the step edge and at the dip ($270^{\circ}$ edge) just below the step edge, while does not appear on the opposite surface of the step edge.
It is also found that the dispersion of the 1D helical conducting states appearing at the step edge and at the dip just below the step edge is linear and gapless, except near the zero energy where a small gap $\Delta E$ is present in their dispersion.
As shall be explained In Sec.~\ref{Sec-Understanding}, the 1D helical conducting state at the dip just below the step edge also originates from the ordinary 1D topological hinge state.
We note that such a linearity and gaplessness feature of the dispersion has already been observed in Fig.~\ref{Fig3}, which comes from the large difference between $N_{c1}$ and $N_{c2}$ i.e., that $N_{c1}\gg N_{c2}$.
On the other hand, we also note that there can be a competition between the increasing number of $N_{c1}$ (or $N_{c2}$) and the bulk bandgap closing, depending on the Hamiltonian parameters.
Indeed, the energy gap near the zero energy $\Delta E$ in the dispersion of the emergent 1D helical conducting states closes in the limit of $(N_{c1}-N_{c2})\to \infty$ as can be seen from Fig.~\ref{Fig6}(b).
In this limit, as is also shown in Figs.~\ref{Fig6}(b) and \ref{Fig6}(d), the dispersion of the three localized states at the leftmost-lower edge, at the rightmost-upper edge, and at the step edge is almost degenerate, which implies that the origin of the localized state at the step edge is the ordinary 1D topological hinge state.
(Note that the incompleteness of the degeneracy of the three states comes from the inversion symmetry breaking nature of the geometry.)
Finally, we show the step height ($\Delta N_{c}=N_{c1}-N_{c2}$) dependence on the energy gap $\Delta E$ in Fig.~\ref{Fig6}(e).
Here, the values of $\Delta E$ at $\Delta N_{c}=10$ and $20$ are those of Figs.~\ref{Fig6}(a) and \ref{Fig6}(b), respectively.

Next, among those parameters, we investigate the effects of $v_b$ and $\lambda_b$ on the electronic structure, which correspond to the orbital-dependent nearest-neighbor electron hopping strengths in the $b$ direction:
\begin{align}
&\sum_{k_b}\psi^\dag_{\bm{k}}\left(v_b\cos k_b \tau_z+\lambda_b\sin k_b\mu_y\tau_y\right) \psi_{\bm{k}}\nonumber\\
&=
\frac{1}{2}\sum_{\langle i_b,i_{b'} \rangle}\left[\psi^\dag_{k_a,k_c}(i_b)\left(v_b\tau_z-i\lambda_b\mu_y\tau_y\right) \psi_{k_a,k_c}(i_{b'})+\mathrm{H.c.}\right],
\end{align}
where $\psi_{\bm{k}}$ is an eight-component electron annihilation operator and $\langle i_b,i_{b'} \rangle$ represents the pairs of nearest-neighbor lattice sites in the $b$ direction.
Since the case of small $\lambda_b$ describes the model we have studied ($\lambda_b=0.1$) in Sec.~\ref{Sec-NumericalResults}, in the following we focus on the effect of $v_b$ with $\lambda_b$ fixed to $\lambda_b=1$.

As shown in Figs.~\ref{Fig7}(a)-\ref{Fig7}(c) and Figs.~\ref{Fig7}(d)-\ref{Fig7}(f), we find that the value of $v_b$ (and $\lambda_b$) plays an essential role in the emergence or absence of the 1D helical conducting state on the opposite surface of a step edge.
We have checked that the bulk bandgap does not close, i.e., the system is still in the higher-order topological insulator phase, when the value of $v_b$ is varied from 1 to 0.4.
It can be seen that, as the value of $v_b$ decreases [corresponding to from Fig.~\ref{Fig7}(d) to Fig.~\ref{Fig7}(f)], the real-space position of the 1D localized state at the dip ($270^{\circ}$ edge) below the step edge moves continuously toward the opposite surface of the step edge.
In other words, a 1D helical conducting state emerges on the opposite surface of a step edge when the electron hopping in the $b$ direction is weak.

We have checked the influence of the height of the geometry on the emergence or absence of the 1D helical conducting state on the opposite surface of a step edge.
For example, in the case of $N_{c1}=20$ and $N_{c2}=10$ [the same geometry as Figs.~\ref{Fig6}(a) and \ref{Fig6}(c)], the 1D helical conducting state does not emerge on the opposite surface even at small values of $v_b$.
Instead, the bulk bandgap closes at a small value of $v_b$.
This result indicates that the preferable geometry is the geometry with a short height, although the emergence is also dependent on the complex competition between the parameters.

We have also checked what would happen in the geometry with a positive-slope step.
As in the case of the calculation intended for WTe$_2$ [Figs.~\ref{Fig2}(b) and \ref{Fig2}(e)], the 1D helical conducting states do not appear at the step edge, at the dip ($270^{\circ}$ edge) below the step edge, or on the opposite surface of the step edge, regardless of the values of $N_{c1}$ and $N_{c2}$.
We shall show later that such an absence can be naturally explained.

\begin{figure}[!t]
\centering
\includegraphics[width=\columnwidth]{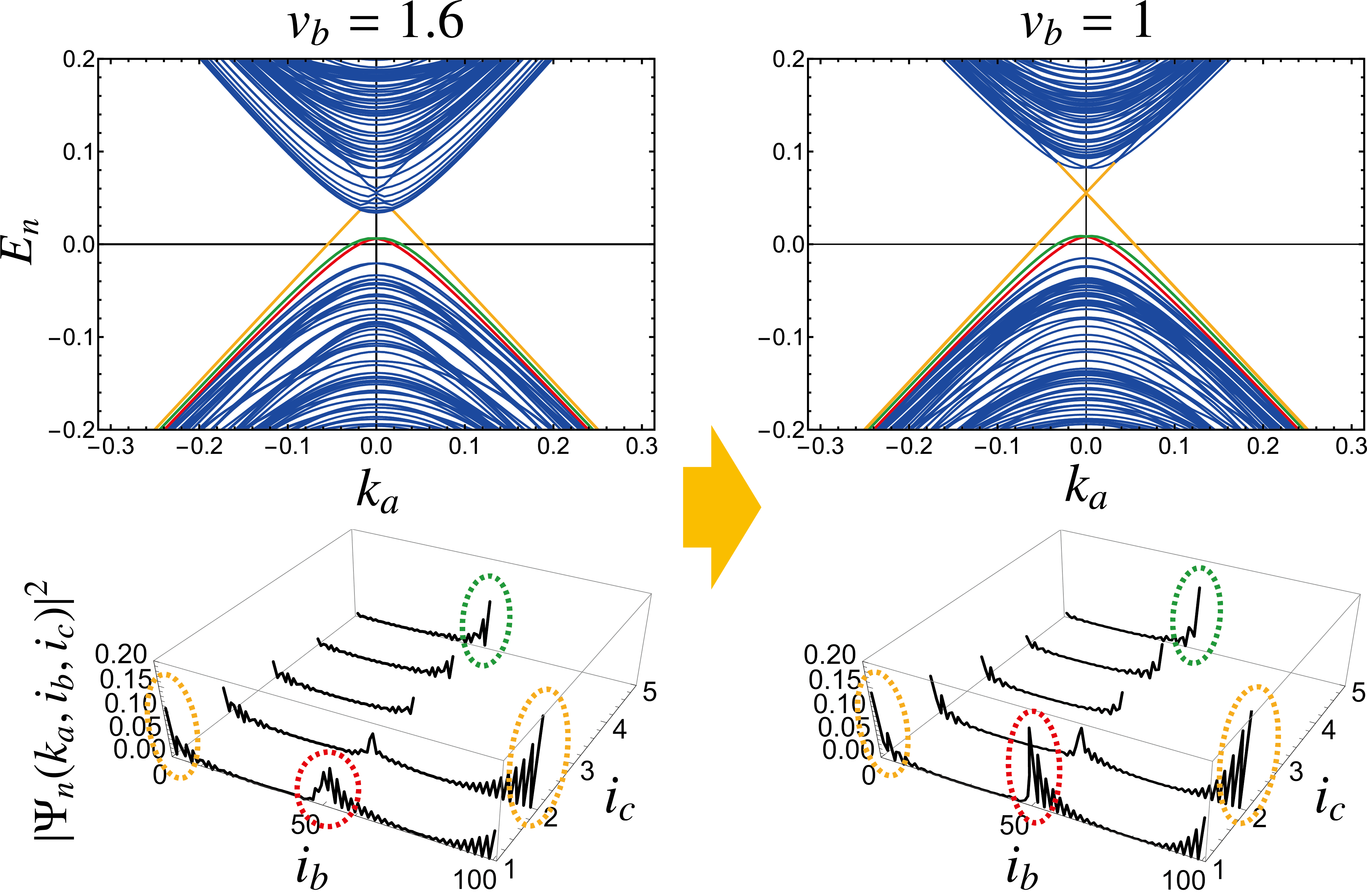}
\caption{
The energy spectrum and site-resolved probability distribution with $k_a=0.05$ of a system with (left) $v_b=1.6$  and (right) $v_b=1$.
The other parameters are the same as in Fig.~\ref{Fig2}.
The system size is $N_{b1}=N_{b2}=50$, $N_{c1}=5$, and $N_{c2}=2$.
}\label{Fig8}
\end{figure}
\begin{figure}[!t]
\centering
\includegraphics[width=\columnwidth]{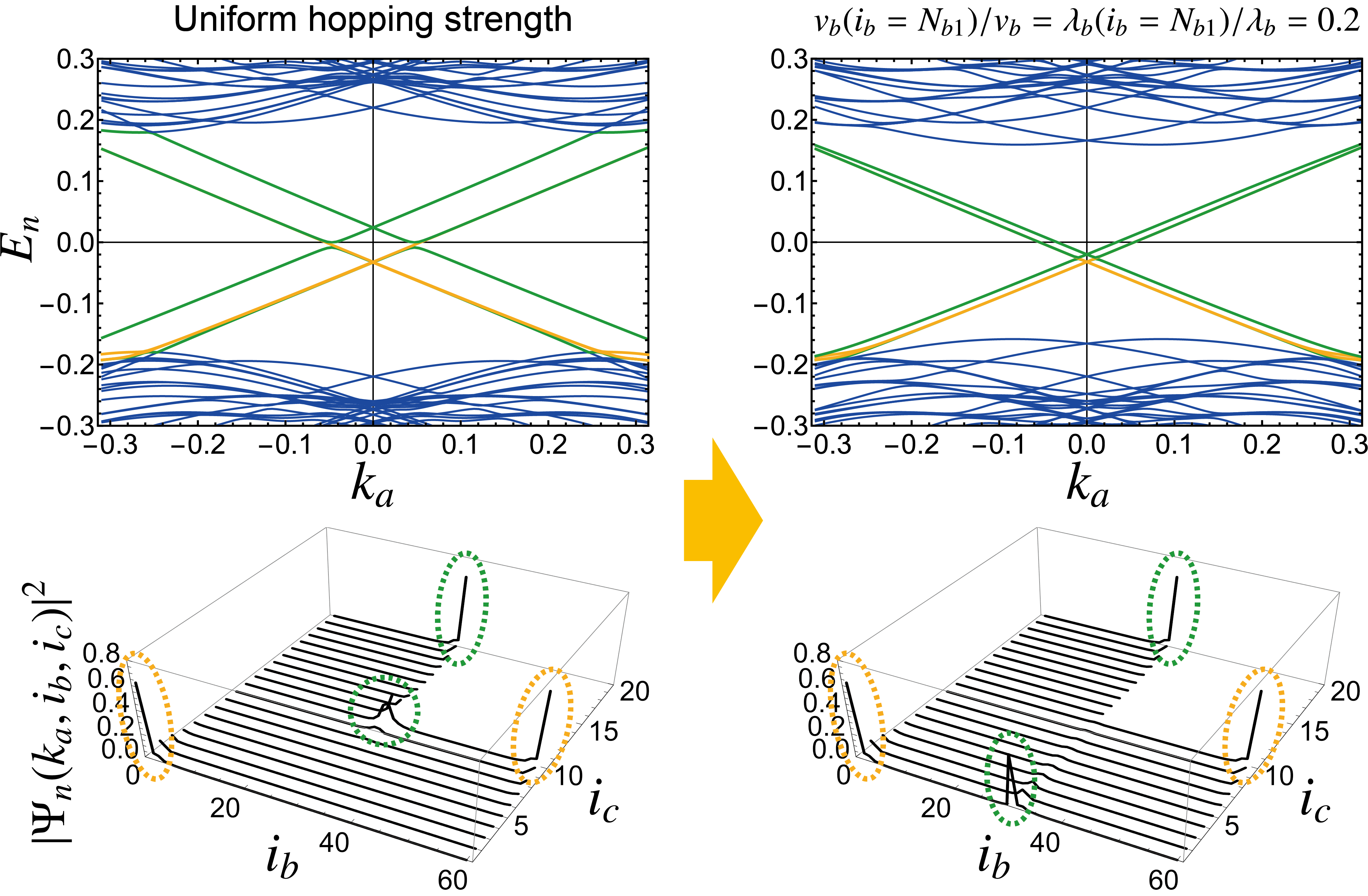}
\caption{The energy spectrum and site-resolved probability distribution with $k_a=0.05$ of a system where (left) the parameters are the same as in Fig.~\ref{Fig6}, and 
(right) the parameters are the same as the left figure except that only the hopping strengths at the boundary of the two blocks are weak such that $v_b(i_b=N_{b1})/v_b=\lambda_b(i_b=N_{b1})/\lambda_b=0.2$.
The system size is $N_{b1}=N_{b2}=30$, $N_{c1}=20$, and $N_{c2}=10$.
}\label{Fig9}
\end{figure}
Finally, we show the result intended for WTe$_2$ in the weak coupling regime in the $b$ direction, i.e., with small values of $v_b$ and $\lambda_b$.
In this case, as we have seen, there exists a 1D helical conducting state on the opposite surface of the step edge even with large values of $v_b$, since the value $\lambda_b=0.1$ is already small enough.
For comparison, we show in Fig.~\ref{Fig8}(left) repeatedly the same energy spectrum and probability distribution as those in Figs.~\ref{Fig2}(c) and \ref{Fig2}(f).
From Fig.~\ref{Fig8}(right) we find that, as the value of $v_b$ becomes smaller, the probability amplitudes of the 1D helical conducting states at the step edge and on the opposite surface of the step edge become larger.
This means that the geometry with a step edge can be regarded as the sum of the two independent blocks  of higher-order topological insulators with height $N_{c1}$ and $N_{c2}$, due to the weak coupling nature of lattice sites in the $b$ direction.

The above physical picture in the weak coupling regime, such that the geometry with a step edge can be regarded as the sum of the two independent blocks, can be further confirmed by considering the case in which only the hopping strengths at the boundary of the two blocks are weak, i.e., only the values of $v_b(i_b=N_{b1})$ and $\lambda_b(i_b=N_{b1})$ are small, compared to the others.
Here, we define $v_b(i_b\neq N_{b1})\equiv v_b$ and $\lambda_b(i_b\neq N_{b1})\equiv \lambda_b$.
Figure~\ref{Fig9} shows the evolution of the electronic structure from the strong coupling case with uniform hopping strengths in the $b$ direction to the case with $v_b(i_b=N_{b1})/v_b=\lambda_b(i_b=N_{b1})/\lambda_b=0.2$.
For comparison, the same energy spectrum and probability distribution as those in Figs.~\ref{Fig6}(a) and \ref{Fig6}(c) are shown repeatedly in Fig.~\ref{Fig9}(left).
We can see from Fig.~\ref{Fig9}(right) that a 1D helical conducting state clearly exists on the opposite surface of the step edge even though the system height is large enough ($N_{c1}=20$ and $N_{c2}=10$).
Moreover, as the values of $v_b(i_b=N_{b1})/v_b$ and $\lambda_b(i_b=N_{b1})/\lambda_b$ become smaller, the overlap of the energy spectrum of the 1D helical conducting states at the step edge and on the opposite surface of the step edge with that of the ordinary 1D topological hinge states becomes stronger.
In the limit of $v_b(i_b=N_{b1})/v_b\to 0$, $\lambda_b(i_b=N_{b1})/\lambda_b\to 0$, the energy spectrum of these four localized states becomes almost degenerate.
Note again that the incompleteness of the degeneracy of the four localized states comes from the inversion symmetry breaking nature of the geometry.
These results also support that the origin of these emergent 1D states is the topological hinge states.
We have checked that the same result, the almost degenerate (as well as gapless and linear) energy spectrum of the four localized states in the limit of $v_b(i_b=N_{b1})/v_b\to 0$, $\lambda_b(i_b=N_{b1})/\lambda_b\to 0$, is also obtained in the case of the parameters intended for WTe$_2$ with $N_{c1}\le 6$.

\subsection{Understanding the emergence \label{Sec-Understanding}}
\begin{figure}[!t]
\centering
\includegraphics[width=\columnwidth]{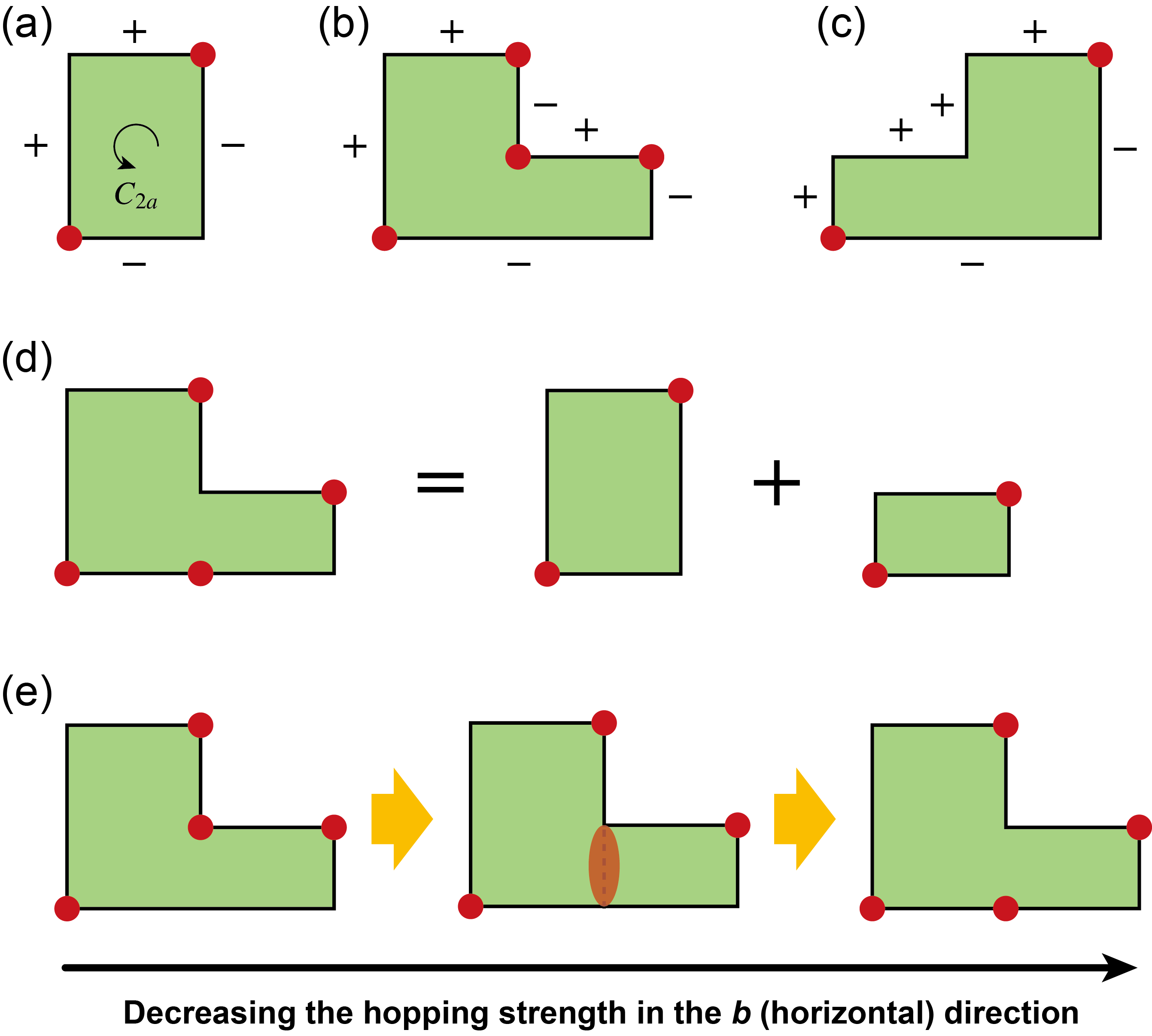}
\caption{
Illustration for understanding the emergence or absence of the 1D helical conducting states in 3D higher-order topological insulators with a step edge.
For concreteness, we here consider the case in which the ordinary 1D topological hinge states appear at the lower-left and upper-right edges of the higher-order topological insulator.
Filled red circles indicate localized 1D helical conducting states.
(a) The ordinary 1D hinge states appear at the corners where the sign ($+$ or $-$) of the mass of the surface Dirac fermions changes.
The system has two-fold rotational symmetry $C_{2a}$ around the $a$ axis and inversion symmetry.
(b) In the geometry with a negative-slope step and a strong electron hopping in the horizontal direction, 1D helical conducting states emerge at the step edge and at the dip ($270^{\circ}$ edge) just below the step edge, in addition to the 1D hinge states at the lower-leftmost and upper-rightmost edges.
(c) In the geometry with a positive-slope step, on the other hand, there only exist the 1D hinge states at the lower-leftmost and upper-rightmost edges.
In (b) and (c), the emergence or absence can be understood by considering the sign change in the mass of surface Dirac fermions.
(d) In a geometry with a negative-slope step and a weak electron hopping in the horizontal direction, a 1D helical conducting state emerges on the opposite surface of the step edge, which can be understood as resulting from an addition of two independent blocks.
(e) Schematic illustration of the evolution of the 1D helical conducting state with respect to the decreasing hopping strength in the horizontal direction.
}\label{Fig10}
\end{figure}
Having the numerical results obtained so far at hand, we are now in a position to discuss the physical mechanism for the emergence of the 1D helical conducting states at a step edge and on the opposite surface of the step edge.

First, we start with a higher-order topological insulator block without a step edge.
For clarity, as we have seen in our calculation for few-layer Td-$X$Te$_2$ ($X$ = Mo, W), we consider the case in which the ordinary 1D topological hinge states appear at the lower-left and upper-right edges of the higher-order topological insulator.
It is known that the ordinary 1D topological hinge states appear at the corners where the sign ($+$ or $-$) of the mass of the surface Dirac fermions changes \cite{Schindler2018,Queiroz2019}.
Thus, in the present case, we may assign for definiteness the left and upper surfaces with the $+$ sign and the right and lower surfaces with the $-$ sign as depicted in Fig.~\ref{Fig10}(a).

Second, we consider the geometry with a negative-slope step.
By applying the same argument as in the case without a step edge, we can see that the 1D helical conducting states emerge where the sign of the mass of the surface Dirac fermions changes, namely, at the step edge and at the dip ($270^{\circ}$ edge) just below the step edge, in addition to the ordinary 1D topological hinge states at the lower-leftmost and upper-rightmost edges [see Fig.~\ref{Fig10}(b)].
In this sense, we argue that these emergent 1D helical conducting states have also a topological origin.
A detailed consideration on the sign assignment to each surface is given in Appendix~\ref{Appendix-Fig10}.
Similarly, we can understand the absence of the 1D helical conducting states at the step edge and at the dip just below the step edge in the geometry with a positive-slope step [see Fig.~\ref{Fig10}(c)].

However, the above consideration based on the sign change in the mass of surface Dirac fermions cannot explain the emergence of the 1D helical conducting state on the opposite surface of a step edge.
Recall here that the emergence on the opposite surface of a step edge is realized when the electron hopping in the horizontal direction ($b$ direction in our case) is weak, as shown in Figs.~\ref{Fig7}(c) and \ref{Fig7}(f).
Note that the result for WTe$_2$ [Fig.~\ref{Fig2}(c) and \ref{Fig2}(f)] is also in the weak coupling regime in the $b$ direction.
In these cases, the emergence on the opposite surface of a step edge can be understood as resulting from an addition of two independent blocks of higher-order topological insulators, as illustrated in Fig.~\ref{Fig10}(d).
This physical picture is directly evidenced from Fig.~\ref{Fig9}(right).
We can see that the helical state with a gapless and linear dispersion survives even when a finite coupling (electron hopping) between the two blocks is present, starting from two decoupled blocks.
This physical picture is also evidenced from that, in the limits such that either width of the two blocks is small enough compared to the other, the larger one can be regarded as a block of a higher-order topological insulator with the 1D topological hinge states at the lower-left and upper right edges, as shown in Fig.~\ref{Fig5} and \ref{Fig-Appendix}.
Interestingly, even in the parameters intended for WTe$_2$, which is already in the weak coupling regime, a further weakening of the hopping strength in the $b$ direction results in an enhancement of the probability amplitude of the 1D helical conducting state on the opposite surface of the step edge (see Fig.~\ref{Fig8}).
This result also supports the physical picture in Fig.~\ref{Fig10}(d).

One may wonder why the weak coupling picture (i.e., addition of two independent blocks) cannot be applied to explain the electronic structure of the geometry with a positive-slope step.
Here, we point out that there is a qualitative difference between the cases of negative-slope and positive-slope steps.
Namely, in the case of the geometry with a negative-slope step, there exists only {\it one} topological hinge state at the boundary between two blocks when they are glued [see the right-hand side of Fig.~\ref{Fig10}(d)], which implies that its presence is stable.
On the other hand,  in the case of the geometry with a positive-slope step, there exist {\it two} topological hinge states at the boundary between two blocks when they are glued [imagine that the large and small blocks are interchanged in Fig.~\ref{Fig10}(d)], which implies that their presence is unstable, i.e., they can be gapped and merge into the bulk states by, so to speak, a pair annihilation.
We think that this difference may lead to the emergence or absence of the 1D helical conducting state on the opposite surface of a step edge, depending on negative-slope and positive-slope steps.

What is common in the above two physical pictures for a system with a step edge is that the dispersion of the 1D helical conducting states emerging at the step edge, at the dip just below the step edge, and on the opposite surface of the step edge becomes linear and gapless near the zero energy in the limit of $N_{c1}\gg N_{c2}$ [see Figs.~\ref{Fig3} and \ref{Fig6}(b)].
This indicates that the origin of these 1D helical conducting states is the ordinary 1D topological hinge states, because the edges where they emerge get to be regarded as independent edges (i.e., the finite-size effect becomes weaker) as the difference $N_{c1}-N_{c2}$ becomes larger.

Summarizing the above, we conclude that (1) the consideration based on the sign change in the mass of surface Dirac fermions can be applied in the strong coupling regime in the $b$ direction, while (2) the consideration based on the combination of two independent blocks of higher-order topological insulators can be applied in the weak coupling regime in the $b$ direction.
As illustrated in Fig.~\ref{Fig10}(e), these two physical pictures are connected continuously without the bulk bandgap closing via a parameter change, i.e., the change in the electron hopping strength in the $b$ direction.
We note that the above argument is naturally extended to the generic case with multiple step edges.

\section{Discussion \label{Sec-Discussion}}
First, let us discuss a possible application of the 1D helical conducting state emerging on the opposite surface of a step edge.
Before that, we note that the Hamiltonian of the emergent 1D helical conducting state around the zero energy can be effectively written as
\begin{align}
H_{\mathrm{1D}}=\int dx\, \psi^\dag (-iv\partial_x\sigma_z-\mu)\psi,
\label{Effective-1D-Hamiltonian}
\end{align}
where $v$ is the slope of the region satisfying the linear dispersion relation $E=vk_a$, $\sigma_z=\mathrm{diag}[1,-1]$ is a Pauli matrix for spin space, and $\psi$ is a two-component spinor.
One of the promising applications of a 1D conducting state is to realize Majorana zero modes \cite{Alicea2012}.
Here, recall a necessary condition of such a 1D conducting state for realizing Majorana zero modes \cite{Alicea2012}: one is a linear dispersion and the other is a helical spin structure, which are both satisfied in Eq.~(\ref{Effective-1D-Hamiltonian}).
Thus, 3D higher-order topological insulators with step edges can be utilized, e.g., as a platform for observing Majorana zero modes.
Although there has already been an experiment suggesting an observation of Majorana zero modes using the ordinary 1D topological helical hinge state of a 3D higher-order topological insulator \cite{Hsu2018,Jack2019}, we here stress that our proposal has an advantage such that one can use the 2D surface (the surface opposite to a step edge) in a transport experiment, making it possible to employ experimental processes commonly used in thin-film devices.

Second, we discuss the relevance of our study to a recent experiment in few-layer WTe$_2$ \cite{Kononov2020}.
As shown in Figs.~\ref{Fig2}(c) and \ref{Fig2}(f), we have found that a 1D helical conducting state emerges at a step edge.
This result is consistent with the experiment, which shows a signature of 1D electronic transport originating from the conducting channels localized at the step edges in few-layer WTe$_2$ through the Josephson effect.
We here stress that the emergence or absence of the 1D helical conducting state at the step edges depends on the structure of the steps with respect to the crystal axes, i.e., the positive- or negative-slope steps we have defined.

Third, we mention a series of theoretical studies on gapless helical modes localized at topological defects including edge and screw dislocations, in view of the fact that a step edge is an edge dislocation.
For step edges characterized by a Burgers vector, the presence or absence of gapless helical modes localized at the step edges can be deduced from the weak topological indices \cite{Ran2009,Teo2010}.
Recently, such a discussion was extended to include higher-order topological insulators \cite{Queiroz2019}. In particular, it was argued that step edges that are not characterized by an integer Burgers vector, but are characterized by the so-called partial dislocation, also host gapless helical modes.
Notice that, in both positive-slope and negative-slope steps we have defined, the Burgers vector is perpendicular to the steps.
In other words, the Burgers vector cannot distinguish the positive-slope steps [where the 1D helical modes are absent at the step edge; see Figs.~\ref{Fig2}(b) and \ref{Fig2}(e)] from negative-slope steps [where the 1D helical modes are present at the step edge; see Figs.~\ref{Fig2}(c) and \ref{Fig2}(f)].
Moreover, the unit cell of Td-$X$Te$_2$ ($X$ = Mo, W) has two layers, and Td-$X$Te$_2$ has trivial $\mathbb{Z}_2$ weak topological indices \cite{Wang2019}.
Thus, the height of the step $|N_{c1}-N_{c2}|$ in our model corresponds to even numbers of layers, which means that such a step is not a partial edge dislocation (but may be regarded as a full edge dislocation).
From these considerations, we think that the mechanism of the emergence of the 1D helical modes at a step edge in our study is different from the mechanism in the above literature.

Fourth, we mention the experimental studies showing the presence of 1D helical conducting states at step edges.
So far, the utilization of step edges has been done mainly in search for large-gap 2D topological insulator (quantum spin Hall insulator) states in layered materials \cite{Drozdov2014,Pauly2015,Li2016,Wu2016,Peng2017,Jia2017,Ohfuchi2022}.
In weak 3D topological insulators Bi$_{14}$Rh$_3$I$_9$ \cite{Pauly2015} and ZrTe$_5$ \cite{Li2016,Wu2016}, a monolayer step edge has been utilized to realize a 1D topological edge channel from the (monolayer) quantum spin Hall state, taking advantage of the nature of weak 3D topological insulators such that weak 3D topological insulators are built by stacking quantum spin Hall insulators.
1D topological edge states have also been observed in 1T$^\prime$-WTe$_2$ with a monolayer step edge \cite{Peng2017,Jia2017}, which, however, are based on that the monolayer 1T$^\prime$-WTe$_2$ is a quantum spin Hall insulator.
Another study has shown the presence of a 1D spin-polarized conducting state at a step edge in a 3D topological crystalline insulator (Pb,Sn)Se \cite{Sessi2016}, where the step edge can be regarded as a domain wall (from the periodicity difference in the crystal structure) of the surface state, and therefore the physical mechanism is similar to that in the Su-Schrieffer-Heeger model \cite{Sessi2016}.
In contrast, the 1D helical conducting states at a step edge and on the opposite surface of the step edge found in our study originate from the 1D topological hinge states.
Namely, their presence is ascribed to the higher-order topology, rather than the quantum spin Hall insulator state or the weak topological insulator state, since few-layer $X$Te$_2$ ($X$ = Mo, W) has trivial $\mathbb{Z}_2$ weak topological indices and is characterized by a $\mathbb{Z}_4$ index \cite{Wang2019}.

Finally, we briefly discuss the stability of the emergent 1D helical conducting states at a step edge and on the opposite surface of the step edge, which is an important future subject that should be elucidated.
Note that their gapped dispersion has been observed when the step height $\Delta N_{c}=N_{c1}-N_{c2}$ is not so large, which means that it is due to the finite-size effect in the $c$ direction [see Fig~\ref{Fig6}(e)].
In other words, the overlap between their wave functions gives rise to an energy gap, as in the case of the edge states of 2D quantum spin Hall insulator strips \cite{Zhou2008}.
Generally, when such an overlap exists, the electron scattering between the state at a step edge and the one on the opposite surface of the step edge can occur.
However, as the finite-size effect becomes weaker, i.e., as the value of $\Delta N_{c}$ becomes larger [as in the cases of Figs.~\ref{Fig3}(c) and \ref{Fig6}(b)], we expect that the emergent 1D helical conducting states become more robust against weak perturbations since the elastic backscattering will be suppressed, and that their robustness becomes similar to that of topological 1D edge or hinge states protected by time-reversal symmetry.
Even within the present study, an implication for the robustness of the helical conducting state on the opposite surface of a step edge might be found in Fig.~\ref{Fig9}.
This figure shows that the helical state with a gapless and linear dispersion survives even when a finite coupling (electron hopping) between the nearest sites is present.
As we have argued, this state originates from the ordinary topological hinge state at the lower-leftmost hinge of the small right-hand-side block.
Although such a finite coupling is not a disorder, the survival of the gaplessness and linearity implies a topological origin of the helical conducting state on the opposite surface of a step edge.

\section{Possible experimental realization and application \label{Sec-Experimental}}
\begin{figure}[!t]
\centering
\includegraphics[width=\columnwidth]{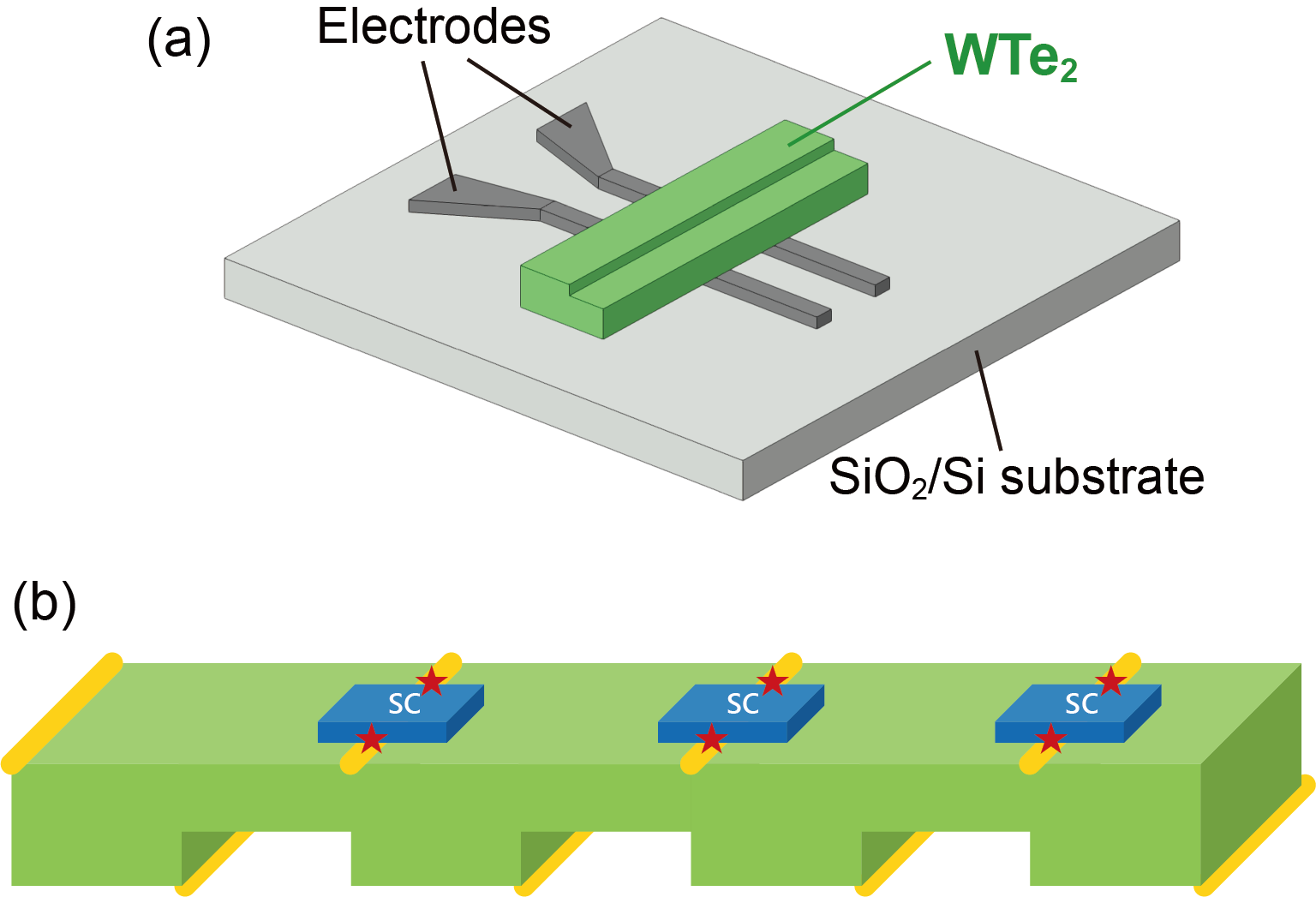}
\caption{(a) Proposed experimental setup to observe the emergent 1D helical conducting channel on the opposite surface of a step edge.
(b) A possible application of the emergent 1D helical conducting channel.
On-demand creation of multiple Majorana fermions (denoted by red stars) in a single material could be possible by creating multiple step edges.
Here, a yellow line and ``SC'' indicate a 1D helical conducting channel and an $s$-wave superconductor, respectively.
}\label{Fig11}
\end{figure}
Here, we propose an experimental process to create a step edge and observe the emergent helical conducting state on the opposite surface of the step edge.
Figure \ref{Fig11}(a) illustrates a bottom contact device in which WTe$_2$ single crystal is placed on top of prepatterned electrodes on SiO$_2$/Si substrate.
The WTe$_2$ crystal can be laminated using standard PDMS stamp method as reported previously \cite{Fei2017,Ohtomo2022}.
After lamination, a part of the crystal surface is covered by electron beam resists, which produce a step on the crystal after Ar ion milling.
Any change in the conductivity before and after the creation of a step can be ascribed to the emergent helical conducting state.
We note that multiple step edges can be created in a similar way as above.
Our method has an advantage such that the bottom surface can be (and therefore the 1D helical channel on the opposite surface of a step edge can also be) free from the surface oxidation, paving the way for a possible realization of high-quality 1D helical conducting channels.

As we have briefly discussed in Sec.~\ref{Sec-Discussion}, it is known theoretically that a 1D helical conducting channel in proximity to an $s$-wave superconductor can realize Majorana zero modes.
Making use of this fact, we expect that on-demand creation of multiple Majorana fermions in a single material could be possible by creating multiple step edges, as illustrated in Fig.~\ref{Fig11}(b).
This on-demand creation might lead to a realization of high-density Majorana qubits.
An advantage of our proposal is that one can use the 2D surface (the surface opposite to a step edge) in a device, making it possible to employ experimental processes commonly used in thin-film devices.

\section{Summary \label{Sec-Summary}}
To summarize, we have found that a 1D conducting state with a helical spin structure, which also have a linear dispersion near the zero energy, emerge at the step edge and on the opposite surface of the step edge in 3D higher-order topological insulators such as few-layer Td-$X$Te$_2$ ($X$ = Mo, W) with a step edge.
Although we have focused on the case with a single step edge, our results are naturally extended to the generic case with multiple step edges.
As we have discussed above, one or more step edges can be created as desired by controlling their locations and numbers.
Therefore, our finding paves the way for on-demand creation of 1D helical conducting states from 3D higher-order topological insulators employing experimental processes commonly used in thin-film devices, which could lead to, e.g., a realization of high-density Majorana qubits.

In order to obtain gapless and linear dispersion inside the bulk bandgap, the height of a step (i.e., the difference $\Delta N_{c}=N_{c1}-N_{c2}$) should be large (see Fig.~\ref{Fig6}).
Note, however, that there can be a competition between the increasing number of $N_{c1}$ (or $N_{c2}$) and the bulk bandgap closing.
Alternatively, gapless and linear dispersion is also obtained if the electron hopping near the boundary between the two subsystems without step edges of the size $N_{b1}\times N_{c1}$ and $N_{b2}\times N_{c2}$ can be weakened locally (see Fig.~\ref{Fig9}), although this may be a little artificial way.
Such a linearity and gaplessness feature of the emergent 1D helical states at the step edge and on the
opposite surface of the step edge in the case of $N_{c1}\gg N_{c2}$ is ascribed to that the origin of the emergent 1D helical states is the ordinary 1D topological hinge state.
By varying the parameters in the tight-binding Hamiltonian, we have also found that the 1D helical conducting state on the opposite surface of a step edge emerges when the electron hopping in the direction perpendicular to the step is weak.
On the other hand, when the electron hopping in the direction perpendicular to the step is strong, a 1D helical conducting state emerges at the dip ($270^{\circ}$ edge) below the step edge instead of emerging on the opposite surface.
Such an emergence at the dip can be understood in the same way as the ordinary 1D topological hinge states, i.e., by that they appear at the corners where the sign ($+$ or $-$) of the mass of the surface Dirac fermions changes.
As illustrated in Fig.~\ref{Fig10}(e), these two physical pictures are connected continuously via a parameter change without the bulk bandgap closing, i.e., without the change in the bulk topology.
Our argument that the origin of the emergent 1D helical state on the opposite surface of a step edge is the ordinary 1D topological hinge state can also be understood from this consideration on the sign changes of the surface Dirac fermion mass.

\acknowledgements
We would like to thank Shintaro Sato, Yoshiyasu Doi, Junichi Yamaguchi, and Masayuki Hosoda for their advice, encouragement, and support.
\\
\\

\noindent
{\bf AUTHOR DECLARATIONS}

\noindent
{\bf Conflict of Interest}

The authors have no conflicts to disclose.
\\

\noindent
{\bf Author Contributions}\\
{\bf Akihiko Sekine:} Conceptualization (equal); Formal analysis (lead);  Investigation (lead); Validation (equal); Writing/original draft (lead); Writing/review \& editing (lead).
{\bf Manabu Ohtomo:} Conceptualization (equal); Writing/original draft (supporting); Writing/review \& editing (equal).
{\bf Kenichi Kawaguchi:} Conceptualization (equal); Writing/review \& editing (equal).
{\bf Mari Ohfuchi:} Conceptualization (equal); Validation (equal); Writing/review \& editing (equal).
\\

\noindent
{\bf DATA AVAILABILITY}

The data that support the findings of this study are available from the corresponding author upon reasonable request.
\\

\begin{widetext}
\appendix
\section{Continuous Deformation of the Geometry with $N_{b1}\gg N_{b2}$ \label{Appendix-Continuous-Deformation}}
\begin{figure*}[!t]
\centering
\includegraphics[width=0.95\columnwidth]{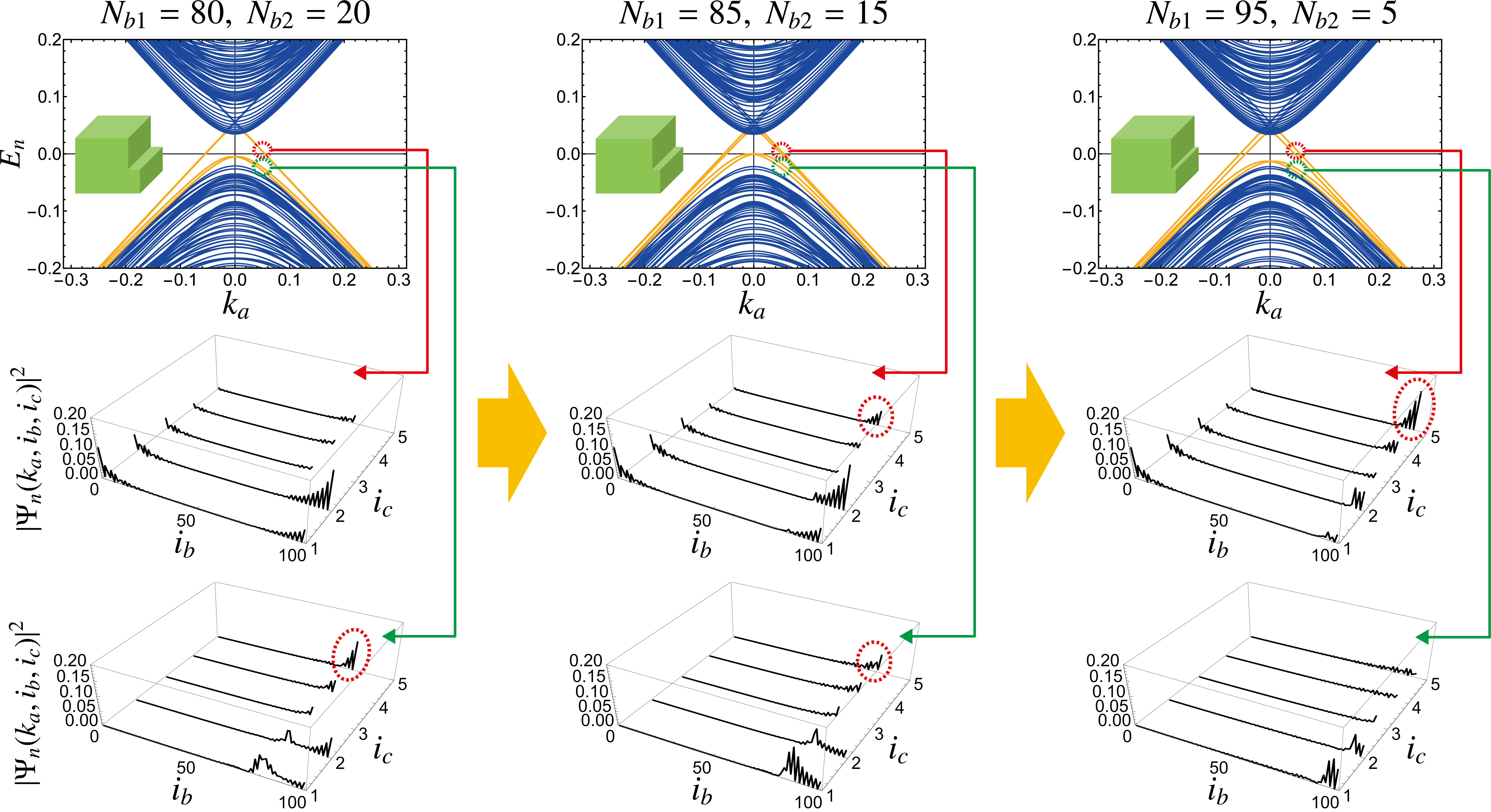}
\caption{Evolution of the energy spectrum and the probability distribution with $k_a=0.05$ under a continuous deformation of the geometry.
From left to right: The width of the left-hand-side block, $N_{b1}$, is varied from $N_{b1}=80$ to $N_{b1}=95$ while the width of the right-hand-side block, $N_{b2}$, is varied under the condition such that $N_{b1}+N_{b2}=100$.
The localized state whose dispersion is gapless (gapped) inside the bulk bandgap is directed by red (green) arrows to the corresponding probability distribution.
Red dashed circles in the probability distributions highlight the 1D localized state at the step edge.
We set $N_{c1}=5$ and $N_{c2}=2$.
The other parameters are the same as in Fig.~\ref{Fig2}.
}\label{Fig-Appendix}
\end{figure*}
In order to understand the origin of the emergent 1D helical conducting state at the step edge, we consider the case in which the width of the left-hand-side block is much larger than that of the right-hand side block (i.e., $N_{b1}\gg N_{b2}$).
Figure~\ref{Fig-Appendix} shows the energy spectrum and probability distribution under a continuous deformation of the geometry with $N_{b1}\gg N_{b2}$.
When the width of the right-hand-side block is quite small ($N_{b2}=5$; right figures), it can be seen that the dispersion of the emergent 1D helical conducting state at the step edge is the same as the dispersion of the ordinary topological hinge states, i.e., is linear and gapless (see the red dashed circle in the probability distribution).
This indicates that the system can be viewed as a single block of a higher-order topological insulator with $N_{c1}=5$.
In other words, the presence of the right-hand-side block does not affect largely the electronic structure of the left-hand-side block.
On the other hand, when the width of the right-hand-side block becomes larger ($N_{b2}=20$; left figures), it can be seen that the dispersion of the emergent 1D helical conducting state at the step edge becomes gapped and the dispersion that is linear and gapless originates from the ordinary topological hinge states at the lower-leftmost and upper-rightmost corners [as we have also seen in Figs.~\ref{Fig2}(c) and \ref{Fig2}(f)].
In between the above two geometries, when the width of the right-hand-side block is intermediate ($N_{b2}=15$; middle figures), an crossover occurs: The state whose dispersion is linear and gapless is localized {\it both} at the upper-rightmost corner and at the step edge.
In other words, a 1D helical conducting state with a linear and gapless dispersion and the one with a gapped dispersion coexist at the step edge.

\section{Detailed Consideration on Fig.~\ref{Fig10} \label{Appendix-Fig10}}
\subsection{From a phenomenological point of view}
Because of the geometry which breaks translational symmetry (i.e., has open boundary conditions) in the two spatial directions, it is not easy to solve the Schr\"{o}dinger equation analytically in the presence of a step edge.
Hence, we here attempt to understand how the assignment of the sign ($+$ or $-$) of the surface Dirac fermion mass in Figs.~\ref{Fig10}(b) and \ref{Fig10}(c) is made from a phenomenological point of view, i.e., from the  energy spectra and wave functions we have observed.

First, by comparing Fig.~\ref{Fig2}(c) with Fig.~\ref{Fig2}(a), we can see that the dispersion of the 1D localized states appearing at the lower-leftmost and upper-rightmost edges (displayed yellow) is gapless and linear even in the geometry with a negative-slope step, and that their dispersion is almost the same as that of the ordinary 1D topological hinge states in the geometry without a step.
Note that, although Fig.~\ref{Fig2} is for the weak-coupling regime in the $b$ direction, we have also the same result for the strong-coupling regime as shown in Fig.~\ref{Fig6}(a).
This implies that these hinge states are protected effectively (or equivalently, locally) by two-fold rotational symmetry $C_{2a}$ around the $a$ axis and by inversion symmetry, as well as those in the geometry without a step [see Fig.~\ref{Fig-Appendix2}(a)].
From this consideration, as shown in Fig.~\ref{Fig-Appendix2}(a), we may assign each surface of the lower part of the geometry with a sign of the surface Dirac fermion mass in the same manner as in the geometry without a step.

Next, in order to identify the signs assigned to the remaining surfaces, we consider the geometry with a negative-slope step in the limit of $N_{c1}\gg N_{c2}$.
In this limit, as can be seen from Fig.~\ref{Fig6}(b), the dispersion of the 1D localized states appearing at the lower-leftmost edge, the upper-rightmost edge, and the step edge is gapless and linear, and their dispersion is almost degenerate (the same as that of the ordinary 1D topological hinge states in the geometry without a step).
This implies that these 1D localized states are protected effectively (or equivalently, locally) by two-fold rotational symmetry $C_{2a}$ around the $a$ axis and by inversion symmetry [see Fig.~\ref{Fig-Appendix2}(b)].
Thus, the signs of the surface Dirac fermion mass may be assigned as shown in Fig.~\ref{Fig-Appendix2}(b).

Finally, combining Figs.~\ref{Fig-Appendix2}(a) and \ref{Fig-Appendix2}(b), we obtain Fig.~\ref{Fig10}(b).
Once the way of assigning the signs of the surface Dirac fermion mass is clarified, the same argument can be applied to the geometry with a positive-slope step.
Namely, we also obtain Fig.~\ref{Fig10}(c).

\begin{figure}[!t]
\centering
\includegraphics[width=\columnwidth]{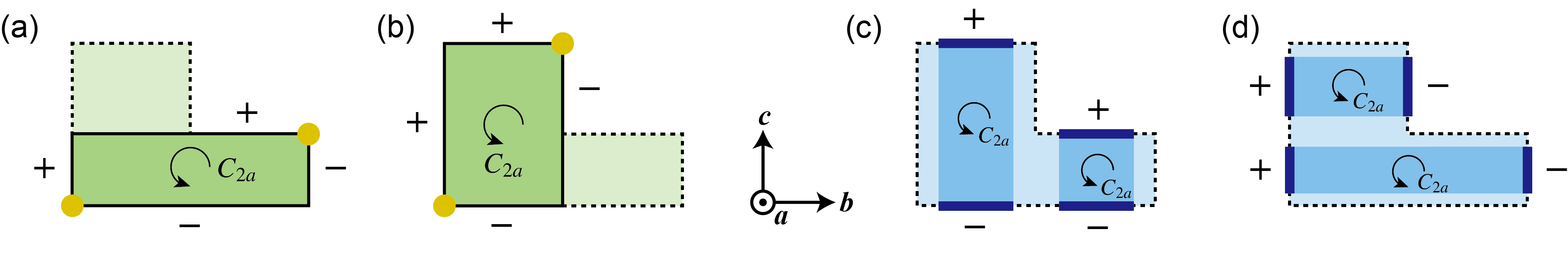}
\caption{(a) The lower part and (b) left part of the geometry with a negative-slope step can be regarded as having effectively (or equivalently, locally) two-fold rotational symmetry $C_{2a}$ around the $a$ axis and inversion symmetry.
Filled yellow circles represent 1D helical conducting states whose dispersion is linear and gapless.
In systems with $N_{b1}\gg 1$,  $N_{b2}\gg 1$, and $N_{c1}\gg N_{c2}$, (c) the left- and right-half parts with surfaces parallel to the $ab$ plane and (d) the upper- and lower-half parts with surfaces parallel to the $ac$ plane can be regarded as subsystems without step edges, since the translation symmetry is locally restored away from the dip ($270^{\circ}$ edge) below the step edge.
}\label{Fig-Appendix2}
\end{figure}
\subsection{From an analytical point of view}
Here, we attempt to understand how the assignment of the sign ($+$ or $-$) of the surface Dirac fermion mass in Figs.~\ref{Fig10}(b) and \ref{Fig10}(c) is made and therefore the resultant emergence of a 1D helical conducting state at the dip ($270^{\circ}$ edge) below a step edge from an analytical point of view based on the Jackiw-Rebbi domain-wall theory.
In the following, we derive the effective Hamiltonian for the surfaces highlighted in Figs.~\ref{Fig-Appendix2}(c) and \ref{Fig-Appendix2}(d), whose interiors can be regarded as subsystems without step edges.
Although the following analysis can be exactly applicable to the geometry without a step edge, it is meaningful to apply the analysis to the geometry with a step edge.
This is because the translation symmetry is locally restored away from the dip, i.e., the subsystems have the same symmetries as the bulk Hamiltonian~(\ref{Effective-Hamiltonian-continuum}), in the limit of $N_{b1}\gg 1$,  $N_{b2}\gg 1$, and $N_{c1}\gg N_{c2}$.

Because of the presence of $C_{2a}$ symmetry and inversion symmetry in the bulk Hamiltonian~(\ref{Effective-Hamiltonian-continuum}), it is sufficient to confirm the sign change in the mass of the surface Dirac fermions for a pair of the surfaces parallel to the $ab$ plane or for a pair of the surfaces parallel to the $ac$ plane.
Below we derive the effective Hamiltonian for the surfaces parallel to the $ab$ plane as illustrated in Fig.~\ref{Fig-Appendix2}(c).

Let us consider a simplified version of the effective Hamiltonian~(\ref{Effective-Hamiltonian-continuum}), which is given by setting $m_2=\gamma_z=0$ in Eq.~(\ref{Effective-Hamiltonian-continuum}) \cite{Ezawa2019}:
\begin{align}
\mathcal{H}_{\mathrm{eff}}(\bm{k})=M(\bm{k})\tau_z+m_3\mu_z\tau_z+\gamma_x\mu_x+\lambda_b\sin k_b\mu_y\tau_y+\lambda_c\sin k_c\tau_x+\beta_a\sin k_a\mu_z\tau_y\sigma_z,
\label{Effective-Hamiltonian-simplified}
\end{align}
where $M(\bm{k})=m_1+\sum_{j=a,b,c}v_j\cos k_j$.
The simplified Hamiltonian~(\ref{Effective-Hamiltonian-simplified}) have the same topological properties as the original Hamiltonian~(\ref{Effective-Hamiltonian-continuum}).
A merit of the simplified Hamiltonian~(\ref{Effective-Hamiltonian-simplified}) is that the system possesses chiral symmetry, and accordingly the band structure is symmetric with respect to the zero energy.
It has been shown that $m_1$ and $v_j$ ($j=a,b,c$) are the parameters that determines the higher-order topology (i.e., a $\mathbb{Z}_4$ index) of the system \cite{Wang2019,Ezawa2019}.
Because the energy spectrum of the gapless hinge modes appear around the $\Gamma$ point in the Brillouin zone, we may consider the continuum Hamiltonian for small $\bm{k}$ in Eq.~(\ref{Effective-Hamiltonian-simplified}).
The existence of the gapless hinge modes implies that the mass $M$ takes a spatially-varying domain wall structure such that $M<0$ inside the higher-order topological insulator and $M>0$ outside of it (i.e.,  in the vacuum or the normal insulator) \cite{Wang2019}.

We derive the effective Hamiltonian for the surfaces normal to the $c$ axis, following Refs.~\cite{Queiroz2019,Queiroz2019a}.
For concreteness, let us suppose that the two surfaces are located at $c=0$ and $c=L$, i.e., suppose a higher-order topological insulator phase in the region $0\le c \le L$ and a vacuum in the regions $c< 0$ and $L<c$.
When the two surfaces are well separated spatially, we may consider the two surfaces normal to the $c$ axis independently, which we call the $c\to 0^+$ and $c\to 0^-$ surfaces.
First, we consider the $c\to 0^+$ surface that is located at $c=0$.
Namely, we require that the system be topologically nontrivial (trivial) when $c>0$ ($c<0$). 
We solve the Schr\"{o}dinger equation $\mathcal{H}_{\mathrm{eff}}(c, k_a, k_b)\psi(c, k_a, k_b)=E\psi(c, k_a, k_b)$, where
\begin{align}
\mathcal{H}_{\mathrm{eff}}(c, k_a, k_b)=\left[M(c)\tau_z-i\lambda_c\partial_c\tau_x\right]+m_3\mu_z\tau_z+\gamma_x\mu_x+\lambda_b k_b\mu_y\tau_y+\beta_a k_a\mu_z\tau_y\sigma_z,
\label{Effective-Hamiltonian-simplified2}
\end{align}
and
\begin{align}
\psi(c, k_a, k_b)=P\Omega(c)\chi(k_a,k_b) \ \ \ \mathrm{with} \ \ \ P\chi(k_a,k_b)=\chi(k_a,k_b).
\end{align}
Here, $M(c)<0$ for $c>0$, $M(c)>0$ for $c<0$, $P$ is the projector which selects the eigenvectors of $\Omega(c)$ that decay exponentially at $c\to \pm\infty$, and $\chi(k_a,k_b)$ is an eigenstate of the projected surface Hamiltonian $\mathcal{H}_{\mathrm{surface}}=P\mathcal{H}_{\mathrm{eff}}P$.
Since the first term in Eq.~(\ref{Effective-Hamiltonian-simplified2}) represents the zero-energy state localized at $c=0$, it follows that
\begin{align}
\left[M(c)\tau_z-i\lambda_c\partial_c\tau_x\right]P\Omega(c)\chi(k_a,k_b)=0.
\label{Differential-Eq}
\end{align}
As is usually done, we can assume the form $P\Omega(c)\chi(k_a,k_b)=e^{+\int dc M(c)/\lambda_c}\chi(k_a,k_b)$.
Substituting this into Eq.~(\ref{Differential-Eq}), we find that
$\chi\propto \bigl[\begin{smallmatrix}1 \\ -i\end{smallmatrix}\bigl]$, which means that $\tau_y\chi=-\chi$.
Equivalently, we find that $P\equiv P_-=\frac{1}{2}(1-\tau_y)$, where the minus subscript indicates that $P_-$ is the projector onto the negative eigenvalue sector of $\tau_y$.
Note that a projector generally satisfies the relation $P^2=P=P^\dag$.
Finally, the  surface Hamiltonian projected to the $c\to 0^+$ surface, which has a gapped spectrum, is obtained as
\begin{align}
\mathcal{H}_{\mathrm{surface}}^{0+}(k_a,k_b)=P_-\mathcal{H}_{\mathrm{eff}}P_-
=-\beta_a k_a\mu_z\sigma_z-\lambda_b k_b\mu_y+\gamma_x\mu_x,
\label{Surface-Hamiltonian-0+}
\end{align}
where we have used that $P_-\tau_yP_-=-1$, $P_-\tau_xP_-=P_-\tau_zP_-=0$, and $P_-\tau_0P_-=1$.

Next, we consider the $c\to 0^-$ surface that is located at $c=L$.
Since the two surfaces are well separated spatially, the origin of the $c$ axis can be redefined and therefore we may require that the system be topologically nontrivial (trivial) when $c<0$ ($c>0$).
In this case, we can assume the form $\psi(c, k_a, k_b)=P\Omega(c)\chi(k_a,k_b)=e^{-\int dc M(c)/\lambda_c}\chi(k_a,k_b)$, where $M(c)<0$ for $c<0$ and $M(c)>0$ for $c>0$.
Substituting this into Eq.~(\ref{Differential-Eq}), we find that
$\chi\propto \bigl[\begin{smallmatrix}1 \\ i\end{smallmatrix}\bigl]$, which means that $\tau_y\chi=+\chi$.
Equivalently, we find that $P\equiv P_+=\frac{1}{2}(1+\tau_y)$, where the plus subscript indicates that $P_+$ is the projector onto the positive eigenvalue sector of $\tau_y$.
Finally, the  surface Hamiltonian projected to the $c\to 0^-$ surface, which has a gapped spectrum, is obtained as
\begin{align}
\mathcal{H}_{\mathrm{surface}}^{0-}(k_a,k_b)=P_+\mathcal{H}_{\mathrm{eff}}P_+
=\beta_a k_a\mu_z\sigma_z+\lambda_b k_b\mu_y+\gamma_x\mu_x,
\label{Surface-Hamiltonian-0-}
\end{align}
where we have used that $P_+\tau_yP_+=1$, $P_+\tau_xP_+=P_+\tau_zP_+=0$, and $P_+\tau_0P_+=1$.
Because the eigenvalues of the Hamiltonian~(\ref{Surface-Hamiltonian-0-}) are symmetric with respect to the zero energy, they are also the eigenvalues of $-\mathcal{H}_{\mathrm{surface}}^{0-}(k_a,k_b)$.
This is equivalent to the fact that we can redefine the Dirac gamma matrices $\Gamma_\mu$ satisfying $\{\Gamma_\mu, \Gamma_\nu\}=2\delta_{\mu\nu}$ by $\tilde{\Gamma}_\mu=-\Gamma_\mu$.
Thus, we can rewrite Eq.~(\ref{Surface-Hamiltonian-0-}) as
\begin{align}
\mathcal{H}_{\mathrm{surface}}^{0-}(k_a,k_b)=-\beta_a k_a\mu_z\sigma_z-\lambda_b k_b\mu_y-\gamma_x\mu_x,
\label{Surface-Hamiltonian-0-2}
\end{align}
which differs from Eq.~(\ref{Surface-Hamiltonian-0+}) only in that the sign of the mass is opposite to each other.
Therefore, we arrive at the situation in Fig.~\ref{Fig-Appendix2}(c).

Finally, combining Figs.~\ref{Fig-Appendix2}(c) and \ref{Fig-Appendix2}(d), we obtain Fig.~\ref{Fig10}(b).
Once the way of assigning the signs of the surface Dirac fermion mass is clarified, the same argument can be applied to the geometry with a positive-slope step.
Namely, we also obtain Fig.~\ref{Fig10}(c).
\end{widetext}


\end{document}